\documentclass[12pt]{article}

\usepackage{xcolor}         
\usepackage{latexsym}
\usepackage{amsmath}
\usepackage{amssymb}
\usepackage{times}
\usepackage{graphicx}
\usepackage{epsfig}
\usepackage{array}
\usepackage{flafter}
\usepackage[page]{appendix}
\usepackage{listings}
\usepackage{algorithm,algpseudocode}
\usepackage{ulem}
\usepackage{verbatim}
\usepackage{hyperref}
\usepackage{qtree}
\usepackage{mathrsfs,amsmath} 
\usepackage{subcaption}
\usepackage{graphicx}

\usepackage[a4paper, total={6in, 8.5in}]{geometry}

\graphicspath{ {Figures/} }

\newcommand{\Rho}{\mathrm{P}}

\newcommand{\bx}{\mathbf{x}}

\newcommand{\Ex}[1]{Example~\ref{ex:#1}}

\newtheorem{definition}{Definition}
\newtheorem{theorem}{Theorem}
\newtheorem{lemma}{Lemma}
\newtheorem{ExampleDef}{Example}

\newcommand{\Example}[3]{
  \begin{list}{}{
      \setlength{\leftmargin}{1em}} 
    \item                           
    \small                          
    \begin{ExampleDef} \rm          
      {\bf \hspace{-1ex}: #1}       
      #2                                
      \hfill {\large \boldmath $\Box$}  
      \label{ex:#3}                      
    \end{ExampleDef}
  \end{list}}

\let\oldthebibliography\thebibliography
\renewcommand\thebibliography[1]{%
  \oldthebibliography{#1}%
  \setlength{\parskip}{0pt}
  \setlength{\itemsep}{0pt}
}

\begin{document}

\begin{center}
{\Large \bf On computing quantum waves exactly \\from classical and relativistic action \par} \vspace{1.5em}

{\large Winfried Lohmiller and Jean-Jacques Slotine \par}
{Nonlinear Systems Laboratory \\
Massachusetts Institute of Technology \\
Cambridge, Massachusetts, 02139, USA\\
{\sl wslohmil@mit.edu, jjs@mit.edu} \par}
\end{center}


\begin{abstract}
We show that the Schr\"odinger equation can be solved exactly based only
on classical least action. Fundamental postulates of quantum mechanics can in turn be derived directly from this construction. The results extend to the relativistic Klein-Gordon, Pauli,  
Dirac, and Maxwell equations, and suggest a smooth transition between physics across scales.

Most quantum mechanics problems have classical versions which involve multiple least action solutions. The associated classical multipaths stem either  from the initial position or momentum distribution, or from branch points, generated, e.g., by a multiply connected manifold (double slit experiment), by spatial inequality constraints (particle in a box), or by a singularity (Coulomb potential).

We show that the exact Schr\"odinger wave function $\psi$ of the original quantum problem can be constructed by combining this classical multi-valued action $\phi$ with the density $\rho$ of the classical position dynamics, where a key point is that $\rho$ can be easily computed from $\phi$ along each extremal action path. The construction is general and does not involve  any quasi-classical approximation.

Examples illustrate how the quantum wave functions for the double-slit experiment or e.g., the hydrogen atom can be computed exactly from their classical least action counterparts. In a quantum measurement process, randomness originates from the determined forward mapping of an initial classical density distribution.
In the Einstein-Podolsky-Rosen experiment, while Bell's inequalities are violated, from this perspective there is indeed a hidden variable in the form of a complex spinor.

These results also provide a simpler computational alternative to Feynman path integrals, as they use only a minimal subset of classical paths and avoid zig-zag paths and time-slicing altogether.
\end{abstract}





\section{Introduction}

Attempts to bridge the conceptual gap between classical and quantum physics have a long and very distinguished history. Central among those is the path-integral formulation of quantum mechanics, starting with Wiener's work on stochastic processes, Dirac's discussion of the relation of classical least action to quantum mechanics~\cite{Dirac33}, Feynman's fundamental paper \cite{Feynman48,Feynman} on path integral computation, and more recent important extensions such as Duru and Kleinert's time reparametrization \cite{Duru, Kleinert2009}. Related work includes hydrodynamic~\cite{Madelung1927} and pilot wave~\cite{Bohm1952} interpretations,  quasi- and semi-classical approximations~\cite{landau_quantum, VanVleck1928, Maslov1972, BerryMount1972Semiclassical}, as well as expansions of quantum commutators based on classical Poisson brackets~\cite{Groenewold1946,moyal1949quantum}.  This paper stems from the same general motivation, but aims to create an {\it exact} and practical construction of Schr\"odinger's wave function based solely on  classical action. 

Section \ref{col} analyzes when classical action solutions of the Lagrangian optimization~\cite{Lagrange}  are multi-valued, due to multiple initial conditions or due to branch points of the associated Hamilton-Jacobi p.d.e.
Branch points can arise, e.g., from multiply connected manifolds, spatial inequality constraints, or singularities of the Hamiltonian. In the double slit experiment, for instance, the two-connected manifold implies a two-valued action solution corresponding to the two shortest connections through the slits. Similarly, for a particle in a box, multiple reflections on the walls with different initial velocities induce multiple local minima of the action. 


Recall that the classical motion of a physical system corresponds to a local extremum over variational paths ${\bf x}(t) = (x^1, ..., x^N) \in \mathbb{R}^N$ of the system's action,  
\begin{equation}
   \phi({\bf x}, t, {\bf x}_o \veebar {\bf p}_o)  \ =  \ \underset{{\bf x}(t)}{\rm{extremum}} 
   \int L(\dot{\bf x}(t), {\bf x}(t),t) \ d t, \ \ \ L \ = \ \frac{1}{2} \ \dot{\bf x}^T {\bf M} \dot{\bf x} +  {\bf A}^T {\bf Q} \ \dot{\bf x}  - V 
   \label{eq:action}  
\end{equation}
with final position ${\bf x}$ at time $t$, initial position ${\bf x}_o$ or (exclusive $\veebar$) initial momentum ${\bf p}_o \ $,  Lagrangian $L$, inertia tensor or metric ${\bf M}({\bf x})$, potential energy $V({\bf x}, t)$, vector potential ${\bf A}({\bf x}, t)$, diagonal charge ${\bf Q}$ matrix with the constant charge $Q$ of each particle on the diagonal, see e.g. \cite{Feynman, Lagrange, Lovelock}. The extremal action field $ \ \phi({\bf x}, t)$ can be computed from the Hamilton-Jacobi p.d.e. \cite{Goldstein, Hamilton, Jacobi}, with $ H \ $  piecewise  ${\bf C}^2({\bf x},t, \nabla \phi)$,
\begin{eqnarray}
- {\frac {\partial \phi}{\partial t}}  =  H  =  {\frac {1}{2}} \ \left( \nabla \phi- {\bf Q} \ {\bf A} \right)^T \ {\bf M}^{-1} \ \left( \nabla \phi-  {\bf Q} \ {\bf A}  \right) + V \label{eq:Hamiltonian}
\end{eqnarray}
In this paper, unless otherwise specified, we will simply use the term action to refer to such local extremal action (or local stationary action). 

The symmetric metric ${\bf M}({\bf x})$ is required to be uniformly invertible, but is not necessarily positive 
definite. We use the standard Laplace-Beltrami tensor operator ~\cite{beltrami1902ricerche, laplace1799mecanique, Lovelock}
\begin{eqnarray}
\nabla_{\bf M} \cdot {\bf f} &=& {\frac {1}{\sqrt {\det {\bf M} }}} \sum_{n=1}^N \frac{\partial }{\partial x^n} \left({\sqrt {\det {\bf M} }} \  f^n  \right) \ \ \ \ {\rm for }\ {\bf f}({\bf x}, t) = (f^1, ..., f^N) \in \mathbb{R}^N  \nonumber \\
\Delta_{\bf M} \ f &=& \nabla_{\bf M} \cdot \left(  {\bf M}^{-1} \nabla f \right),  \ 
\nabla f = \frac{\partial f}{\partial {\bf x}} \ \ \ \ \ \ \ \ \ \ \ {\rm for }\ f({\bf x}, t) \in \mathbb{R} \label{eq:Delta} 
\end{eqnarray}
for a given metric ${\bf M}({\bf x})$. No index is used for ${\bf M}({\bf x}) = {\bf I}$. The vector potential ${\bf A}({\bf x},t)$ is assumed to follow the Coulomb or Lorenz gauge $\nabla_{\bf M} \cdot {\bf A} = 0$ \cite{Lorenz1867, feynman_quantum_1998}. 

Section \ref{quan} shows that this multi-valued action can be converted exactly into the quantum wave function, provided it is weighted along each branch according to the classical {\it density} of the velocity field implied by the action. A key point is that the classical density $ \ \rho \ $ can be computed analytically along each action branch simply by using the classical continuity equation \cite{Euler},
\begin{equation}
    0 \ = \ \frac{\partial}{\partial t} \ \rho + \nabla_{\bf M} \cdot (\rho \ {\dot{\bf x}} ) \ = \
\frac{d \rho }{dt} \ + \rho \ \nabla_{\bf M} \cdot {\dot{\bf x}}   \label{eq:continuity}
\end{equation}
We will see that this evolution of the classical density distribution over time implies the time evolution of the quantum probability density distribution over time. 

The above construction is general and does not involve  any quasi- or semi-classical approximation.
Dirac \cite{Dirac33} introduced an approximate relation between the single-valued action (\ref{eq:action}) and the quantum wave $\psi$ of the Schr\"odinger equation,
\begin{equation}
    \psi \approx e^{\frac{i }{\hbar} \phi ({\bf x},t) } \label{eq:Dirac}
\end{equation}
under the assumption $ \ \hbar \ \Delta_{\bf M} \phi \approx 0$, with $\hbar$ the reduced Planck constant \cite{planck1901theorie}. Under the same assumption, more general quasi-classical expansions \cite{landau_quantum} later used the 1-dimensional approximation $ \ \psi \approx \frac{C_1}{\sqrt{|\nabla \phi}|} \ e^{\frac{i }{\hbar} \phi} + \frac{C_2}{\sqrt{|\nabla \phi|}} \ e^{-\frac{i }{\hbar} \phi \ }$ where $C_1, C_2$ are constants. Another semi-classical approximation,  $ \ \psi({\bf x}, {\bf x}_o, t) \approx \sum_{class. \ paths\ j \ } (2 \pi i \hbar)^{-\frac{N}{2}} \ \left| \det (\frac{\partial^2 \phi_j}{\partial {\bf x} \partial {\bf x}_o} ) \right|^{\frac{1}{2}} \ e^{\frac{i }{\hbar} \phi_j - \frac{i \pi}{2} \nu_j} \ $,  was derived from the Van Vleck determinant~\cite{VanVleck1928} an the associated Maslov indices \cite{Maslov1972}, which count the number of times $\nu_j$ each classical path crosses a caustic and introduces accordingly a phase correction $ - \frac{\pi}{2}\ \nu_j \ $. 
All semi-classical approximations become quite incorrect for small masses $M$ close to constraints or to singular potential fields, where actually $ \ \Delta_{\bf M} \phi \ $ 
can become unbounded, reflecting the rapid change in local momentum. 

As detailed in Section~\ref{quan}, semi-classical approximations can actually be replaced by an exact result, by substituting 
the full set of extremal actions $\phi_j$ for Dirac's action $\phi$ in (\ref{eq:Dirac}), and {\it weighting} each resulting term
$ \ e^{\frac{i }{\hbar} \phi_j} \ $  by the square root of the classical density $ \rho_j \ $, computed from (\ref{eq:continuity}) along that action branch. In other words, Feynman's infinity of non-classical zig-zag paths can be reduced to the subset of all extremal classical action paths, each weighted according to the classical fluid density computed along the path. This yields an exact construction of the wave function $\psi ({\bf x},t)$ of the Schr\"odinger equation \cite{Cohen-Tannoudji, Schrodinger1926},
\begin{equation}
0 = \left[ \frac{\hbar}{i}  \frac{\partial }{\partial t} + \frac{1}{2}\left( \frac{\hbar}{i} \nabla_{\bf M} - {\bf Q} \ {\bf A} \right) \cdot {\bf M}^{-1} \left( \frac{\hbar}{i} \nabla -  {\bf Q} \ {\bf A} \right) + V  \right]  \psi  \label{eq:Schroed}
\end{equation}
Examples illustrate the computation of the wave function based on the set of all classical action solutions and their associated classical densities. They include the double slit experiment, the Aharonov–Bohm effect, a particle in a box, quantum tunneling, a harmonic oscillator, and the Coulomb potential of a hydrogen atom and its relation to Kepler orbits. We also show how 
wave collapse at measurement \cite{Born1926ZurQD} is a consequence of our formalism.

Section \ref{relativity} extends the results to the relativistic Klein-Gordon \cite{Fulling, Gordon, Klein}, Dirac, Pauli \cite{Pauli, Dirac28} and Maxwell equations \cite{Maxwell1865}. Illustrative examples include the Einstein-Podolsky-Rosen experiment  with entangled spinning particles, and positron-electron creation in quantum electrodynamics.

Concluding remarks and perspectives are offered in Section~\ref{concluding}.

\section{\label{col} Multi-valued local least action and multipaths} 

Rather than being based on an approximation or a mathematical expansion, our mapping from classical to quantum mechanics relies on a different and exact mechanism. This mechanism is built on the combination of an exhaustive census of the multiple classical action solutions, which we first discuss, and of a key algebraic result relating the solutions of the Schr\"odinger equation to those of the Hamilton-Jacobi p.d.e. (Section~\ref{quan}, Lemma~\ref{lem:equivalence}).

Let us first introduce spatial inequality constraints on a multiply connected manifold, and then define action branches on this manifold.
\begin{definition} 
The constrained multiply connected manifold $ \ \mathbb{G}^N \subseteq \mathbb{R}^N$ is defined by the $g=1, ..., G$ inequality constraints $ \ f_g({\bf x}, t)  \le  0 \ $.
The set of active constraints $\mathbb{G}({\bf x}, t) \subseteq  \{1, ..., G \}$  is the set of indices $g$  on the boundary $\partial \mathbb{G}^N$ of $\mathbb{G}^N$ such that $ \ f_g({\bf x}, t) = 0 \ $.

At the time of reflection on $\partial \mathbb{G}^N$, the kinetic energy is preserved for a fully elastic reflection, whereas it
vanishes for a plastic collision.
 \label{def:equalconstraint}
\end{definition}
 
\begin{definition}
The action branch set is the set of local extremal action fields $ \phi_j({\bf x}, t), 
\ {\bf x} \in \mathbb{G}^N, t \ge 0$ which are different  at least at one $({\bf x}, t)  \ $, except for the integration constant for each given initial condition in (\ref{eq:action}).
\label{def:branchset}
\end{definition}
Note that by this definition, a single action branch $ \phi_j({\bf x}, t)$ always spans the complete constrained manifold $\mathbb{G}^N$ even if it is partially identical to another action branch.
Also, omitting a branch $j$ in the calculation of the multi-valued action is equivalent to assuming that the initial density $\rho_j$ of this branch is zero. 

Multi-valued least action branches naturally arise from multiple initial conditions in ${\bf x}_o$ or ${\bf p}_o \ $ in (\ref{eq:action}).  After the initialization, they can also arise at branch points. 
\begin{definition}

The set of branch points $\mathbb{B}^N\subseteq  \mathbb{G}^N$ consists of all points where the set of distinct  $ \ \phi_j({\bf x}, t)$ changes, for $t \ge 0 \ $. Each branch point ${\bf x}$ occurs $(b_{\bf x}-1)$ times in $\mathbb{B}^N$, where $ \ b_{\bf x} \in \mathbb{N} \ $ is the number of distinct action solutions in the local neighborhood of ${\bf x}$.
\label{def:branch}
\end{definition}
The extremization (\ref{eq:action}) and the associated Euler-Lagrange and Hamilton o.d.e. for ${\bf x} \in  \mathbb{R}^N$ can be extended to the case of constrained positions ${\bf x} \in \mathbb{G}^N \subset \mathbb{R}^N$ of Definition~\ref{def:equalconstraint}. The action $\phi$ of (\ref{eq:action}) has a local extremum \cite{Lagrange} if the variation of the action (\ref{eq:action}) 
\begin{eqnarray}
\delta \phi  &=&  \int_o^t  \frac{\partial L}{\partial \dot{\bf x}} \ \delta \dot{\bf x} +  \frac{\partial L}{\partial {\bf x}} \ \delta {\bf x} \ d \theta  \nonumber \\
&=& \left[   \frac{\partial L}{\partial \dot{\bf x}} \ \delta {\bf x}  \right]_o^t - \int_o^t \left( \frac{d}{dt} \frac{\partial L}{\partial \dot{\bf x}} - \frac{\partial L}{\partial {\bf x}}  \right) \delta {\bf x} \ d \theta = \int_o^t \sum_{g \in \mathbb{G}}  \lambda_g \frac{\partial f_g}{\partial {\bf x}} \ \delta {\bf x} \ d\theta \nonumber 
\end{eqnarray}
is only non-zero orthogonal to an active constraint, where the Lagrange parameter $\lambda_g$ defines the magnitude of the cost gradient at the active constraint. The first term on the right-hand side is zero since $\delta {\bf x}$ is zero at the start and end points. In between $\delta {\bf x}$  can take on any arbitrary value. Thus a local least action solution satisfies
\begin{equation}
   \frac{d}{dt} \frac{\partial L}{\partial \dot{\bf x}} - \frac{\partial L}{\partial {\bf x}} \ = \ \frac{d \nabla \phi}{dt} + \frac{\partial H}  {\partial {\bf x}} \ = \ \sum_{g \in \mathbb{G}} \lambda_g \frac{\partial f_g}{\partial {\bf x}} \label{eq:EulerLangrange}
\end{equation}
This extends the usual Euler-Lagrange or Hamilton's o.d.e. (see e.g. \cite{Goldstein, Lagrange}) with Lagrangian collision forces activated by inequality constraints. 

The existence and uniqueness of a solution of (\ref{eq:EulerLangrange}) is guaranteed by the Picard-Lindelöf Theorem \cite{Picard1890} for Lipschitz \cite{Lipschitz1869} continuous $-\frac{\partial H}  {\partial {\bf x}} + \sum_{g \in \mathbb{G}} \lambda_g \frac{\partial f_g}{\partial {\bf x}}$. Contraction theory \cite{lohmiller1998contraction} guarantees a bounded contraction rate of the Hamilton-Jacobi p.d.e. (\ref{eq:Hamiltonian}) for bounded $\Delta_{\bf M} \phi_j({\bf x} \in \mathbb{G}^N, t)$ and $\nabla \phi_j({\bf x} \in \partial \mathbb{G}^N, t)$ \cite{lohmiller2005contraction}. For a given initial condition, a bounded contraction rate in turn implies the existence and uniqueness of a solution of the Hamilton-Jacobi p.d.e. (\ref{eq:Hamiltonian}). Thus, the branch points of Definition \ref{def:branch} can only occur at unbounded $\Delta_{\bf M} \phi_j({\bf x} \in \mathbb{G}^N, t)$ or unbounded $\nabla \phi_j({\bf x} \in \partial \mathbb{G}^N, t)$.
 
These properties and (\ref{eq:continuity}) may be summarized as follows, introducing for generality an ensemble $\mathbb{E}$ of possible initial density conditions to describe different initial conditions occurring with given probabilities.
\begin{theorem}
The  multi-valued least action field $ \ \phi_j({\bf x}, t, {\bf x}_o \veebar {\bf p}_o)$, with $H$ piecewise ${\bf C}^2({\bf x},t, \nabla \phi)$,
\begin{eqnarray}
- {\frac {\partial \phi_j}{\partial t}} \ = \ H \  = \ \frac {1}{2} \ \left( \nabla \phi_j - {\bf Q} \ {\bf A} \right)^T \ {\bf M}^{-1} \ \left( \nabla \phi_j -  {\bf Q} \ {\bf A}  \right) \ + \ V \ \ \ \ \ \ \ \ \ \ \forall t \ge 0  \label{eq:HJ} 
\end{eqnarray}
locally extremizes (\ref{eq:action}) on the constrained multiply connected manifold ${\bf x} \in \mathbb{G}^N$ of Definition \ref{def:equalconstraint}, yielding the multipaths
\begin{eqnarray}
   \frac{d \nabla \phi_j}{dt}  +  \frac{\partial H}  {\partial {\bf x}}\  &=&  \sum_{g \in \mathbb{G}} \frac{\partial f_g}{\partial {\bf x}} \lambda_g \label{eq:dotp} \\
   {\bf M}({\bf x})\ \frac{d{\bf x}}{dt}  &=& \nabla \phi_j- {\bf Q} \ {\bf A} \label{eq:dotq} 
\end{eqnarray}
where the reflection force $\lambda_g \ $ fulfills Definition \ref{def:equalconstraint}. 

The action is said to be $\mathbb{J}$-valued, where $\mathbb{J}$ indexes the set
\begin{equation}
     \{ {\bf x}_o \ \veebar \ {\bf p}_o \} \ \times \ \mathbb{B}^N   \label{eq:branchpointset}
\end{equation}
which accounts for all initial conditions in ${\bf x}_o$ or ${\bf p}_o$ in (\ref{eq:action}) and all branch points $\mathbb{B}^N$ of Definition \ref{def:branch}. Branch points exist only at unbounded  $\Delta_{\bf M} \phi_j({\bf x} \in \mathbb{G}^N, t)$ or unbounded $\nabla \phi_j({\bf x} \in \partial \mathbb{G}^N, t)$. $\mathbb{B}$ indexes all branch points $\mathbb{B}^N$.
Each element of $ \ \mathbb{J} \ $ corresponds to an action branch $\phi_j \ $, while each element of $ \ \mathbb{B} \ $ corresponds to an action branch $\phi_j \ $ for a specific initial condition $\{ {\bf x}_o \ \veebar \ {\bf p}_o \}$.

Combining (\ref{eq:continuity}), (\ref{eq:dotq}) the classical density can be computed along each extremal path ${\bf x}_j(t)$, yielding the path integral 
\begin{equation}
   \rho_j^{\epsilon} ({\bf x}_j(t), t ) = \rho_{oj}^{\epsilon} \ e^{- \int_o^{t} \Delta_{\bf M} \phi_j({\bf x}(\theta), \theta) \ d \theta}  
      \label{eq:density}
\end{equation}
where the gauge $ \ \nabla_{\bf M} \cdot {\bf A} = 0 $ was used. The initial density $\rho_{oj}^{\epsilon} = \rho_j^{\epsilon}({\bf x}_o \veebar {\bf p}_o, 0) $ is normalized to the real probability $ \ \int_{\mathbb{G}^n}  \rho_{oj}^{\epsilon}  \ dx^1 ... d x^N  = \ p^{\epsilon} \ge 0 \ $ of each element $ \ \epsilon \ $ in an ensemble $\mathbb{E} \ $,  with $\sum_{\epsilon \in \mathbb{E}} p^{\epsilon}  = 1$. The density $\rho_j^{\epsilon} ({\bf x}_j(t), t )$ corresponds to the classical probability density to find the Hamiltonian system at ${\bf x}_j$ at time $t$. The superscript ${\epsilon} $ is removed if the ensemble has only one element.
\label{th:Hamilton}
\end{theorem}
Theorem \ref{th:Hamilton} computes the $\mathbb{B}$ multipaths connecting an initial ${\bf x}_o \ \veebar \ {\bf p}_o $ to a final ${\bf x}$. Its result is {\it determined} in that sense, but in contrast to Newton's law \cite{Newton1687} it is in general {\it not deterministic} since the initial condition is incomplete, as initially only ${\bf x}_o$ or ${\bf p}_o$ can be defined. The continuity equation (\ref{eq:density}) is a {\it classical stochastic description} of how an initial probability density distribution propagates in the future, although no noise source exists in the Hamiltonian (\ref{eq:HJ}) beyond the non-Lipschitz constraint force in (\ref{eq:dotp}). Hence the incompleteness of the initial condition leads to a classical stochastic density field $\rho_j({\bf x}_j, t )$ from an initial ${\bf x}_o \ \veebar \ {\bf p}_o $ to any final ${\bf x} \in \mathbb{G}^N$. Finally, the original derivation by Hamilton and Jacobi \cite{Hamilton, Jacobi}, defined on a manifold $\mathbb{R}^N$ rather than on a constrained manifold $\mathbb{G}^N$, was not formulated to predict multipath solutions beyond the multiplicity of initial conditions in (\ref{eq:branchpointset}). We will see that using all these classical features of Theorem \ref{th:Hamilton}, rather then using Newton's single deterministic path \cite{Newton1687}, allows to resolve most conflicts of quantum physics with classical physics.

\section{Exact wave computation from classical multi-valued action and density} \label{quan}

We now build on the above result to show that the Schr\"odinger equation can be solved exactly by computing wave functions directly from the Hamilton-Jacobi p.d.e. A the time of Schr\"odinger and Feynman, the Hamilton-Jacobi p.d.e. only had a single-valued least action $\phi$ for a given initial condition. Hence the Feynman path integral~\cite{Feynman} had to consider all non-classical stochastic zig-zag paths with a time slicing approach, rather than just those minimizing (\ref{eq:action}). This stochastic process noise along the path can be avoided if one uses the deterministic multi-valued local least action solutions of Theorem \ref{th:Hamilton}. As we now show, the Schr\"odinger equation (\ref{eq:Schroed}) can be solved on each extremal branch $j$ of Definition \ref{def:branchset} by 
\begin{equation}
\psi_j = \sqrt{\rho_j} \ e^{\frac{i }{\hbar} \phi_j} = \sqrt{\rho_{oj}}({\bf x}_o \veebar {\bf p}_o, 0)  \ e^{- \frac{1}{2} \int_{0}^{t} \Delta_{\bf M} \phi_j({\bf x}(\theta), \theta) \ d \theta + \frac{i }{\hbar} \phi_j} \label{eq:psij} 
\end{equation}
using the action field $\phi_j({\bf x}, t, {\bf x}_o \veebar {\bf p}_o)$ and the classical density path integral (\ref{eq:density}) of Theorem \ref{th:Hamilton}. Taking the sum of the waves $\psi_j$ over all branches $j$ then yields the overall wave $\psi$. 

\begin{lemma} For each branch $j$, plugging the piecewise ${\bf C}^2$ wave $\psi_j$ from  (\ref{eq:psij}) into the Schr\"odinger equation (\ref{eq:Schroed}) exactly leads to the Hamilton-Jacobi p.d.e. (\ref{eq:HJ}),
 \begin{eqnarray} \label{action_to_wave}
 && \left[ \frac{\hbar}{i}  \ \frac{\partial }{\partial t} + \frac{1}{2}\left( \frac{\hbar}{i} \nabla_{\bf M} -  {\bf Q} \ {\bf A}   \right) \cdot {\bf M}^{-1} \left( \frac{\hbar}{i} \nabla - {\bf Q} \ {\bf A} \right) + V  \right] \ \psi_j 
\nonumber \\
& =& \left[ \frac{\partial \phi_j}{\partial t}  + {\frac {1}{2}} \ \left( \nabla \phi_j- {\bf Q} \ {\bf A} \right)^T \ {\bf M}^{-1} \left( \nabla \phi_j- {\bf Q} \ {\bf A} \right) + V \right] \ \psi_j  \ = \ 0 \label{eq:equivalence}
\end{eqnarray} 
The first equation is an operator equation, which becomes a product in the second equation thanks to the exponential form of the wave $\psi_j$. Equation (\ref{eq:equivalence}) holds for piecewise ${\bf C}^2$ real, complex or quaternion actions and waves.
\label{lem:equivalence}
\end{lemma}

\noindent {\bf Proof} \ \ The density (\ref{eq:density}) is only defined along each individual path ${\bf x}(t)$  (\ref{eq:dotq}), so that its total differential has no variation with respect to ${\bf x}$. Replacing $\psi_j$ in the first line of (\ref{eq:equivalence}) yields
\begin{equation} \left[ \frac{\partial \phi_j}{\partial t} - \frac{\hbar}{2 i}  \Delta_{\bf M} \phi_j  + {\frac {1}{2}} \ \left( \nabla \phi_j- {\bf Q} \  {\bf A} \right)^T \ {\bf M}^{-1} \left( \nabla \phi_j- {\bf Q} \  {\bf A} \right) +\frac{\hbar}{2 i}  \Delta_{\bf M} \phi_j  \right] \ \psi_j \nonumber
\end{equation}
which in turn yields the second line. 
$ \hfill \square$ 

In earlier attempts~\cite{Madelung1927,Schleich2013SchrdingerER, VanVleck1928, Maslov1972} to map the Hamilton-Jacobi equation to the Schr\"odinger equation, (\ref{eq:equivalence}) needs to be extended by an extra error term, since the normalization factor is based on a {\it general field} $R({\bf x}, t)$, not necessarily fulfilling the continuity equation (\ref{eq:continuity}). Instead, for a given ${\bf x}_o$ or ${\bf p}_o$ at $t = 0$, the proposed density path integral $\rho_j({\bf x}(t), t)$ of (\ref{eq:density})  fulfills (\ref{eq:continuity}) and depends only on $t$ along the path ${\bf x}(t)$.  
Note that both classical density and classical action are path integrals, where the density is path-dependent and the action is path-independent. Also, earlier attempts do not always use a complete $\mathbb{J}$-valued action of (\ref{eq:branchpointset}). In this paper we may use interchangeably the notations $\sum$ and $\int$ in normalized sums as some indices may include mixtures of discrete and continuous quantities. 
\begin{theorem} 
The wave function $\psi^{\epsilon}({\bf x}, t)$ of the Schr\"odinger equation  (\ref{eq:Schroed}) can be computed from the multi-valued least action field $\phi_j({\bf x}, t, {\bf x}_o \veebar {\bf p}_o)$, the initial action field $\phi_{oj}({\bf x}, 0, {\bf x}_o \veebar {\bf p}_o)$, the classical density path integral (\ref{eq:density}) of the ensemble $\mathbb{E} \ $ of Theorem \ref{th:Hamilton}
\begin{eqnarray}
  \psi^{\epsilon} &=& \sum_{j \in \mathbb{J}}   \sqrt{\rho_j^{\epsilon}}  \ e^{\frac{i}{\hbar} \phi_j}   \label{eq:Wave} 
\end{eqnarray}
For each $ \ j$, the action provides the phase of the wave and the square root density normalized as above gives the corresponding gain. 
Equivalently,  $\psi^{\epsilon}({\bf x}, t)$ can be expressed using the Feynman kernel $K_j({\bf x}, t, {\bf x}_o \veebar {\bf p}_o)$ \cite{Feynman} computed from classical action and density,
\begin{eqnarray}
  \psi^{\epsilon}  \ = \ \sum_{j \in \mathbb{J}}  
  K_j \ \psi_{oj}^{\epsilon} \ \ \ \ \ \ \rm{where} \ \ \  K_j = \sqrt{ \frac{\rho_j}{\rho_{oj}}} \ e^{\frac{i }{\hbar} (\phi_j-\phi_{oj})} \ \ \ \  \rm{and}\ \ \  \psi_{oj}^{\epsilon} = \sqrt{  \rho_{oj}^{\epsilon}}  e^{\frac{i }{\hbar} \phi_{oj}} \label{eq:Feynman_kernel} 
\end{eqnarray}

The associated  quantum density matrix at time $t$
\begin{equation}
\varrho ({\bf x}, t) = \sum_{\epsilon \in \mathbb{E}} p^{\epsilon} \ \psi^{\epsilon} \psi^{\epsilon \dagger} \ \ \  \label{eq:Prob}
\end{equation}
is the determined forward mapping along all classical paths of Theorem \ref{th:Hamilton} from the initial quantum density distribution at $t=0$. 
\label{th:quantum}
\end{theorem}

Note that an initially normalized distribution (\ref{eq:density}, \ref{eq:Wave}) remains normalized $\ \forall t \ge 0$, as (\ref{eq:Schroed}) 
and its conjugate imply that
\begin{equation}
    \frac{\partial }{\partial t} \int_{\mathbb{G}^n}   \psi^{\epsilon \dagger}\psi^{\epsilon} \ dx^1 ... d x^N = 0 \label{eq:prob1}
\end{equation}
Also note that Theorems \ref{th:Hamilton} and \ref{th:quantum} immediately extend 
to complex or quaternion actions, as long as a real classical path can be constructed, for instance using superposition. Equation (\ref{eq:Wave}) of Theorem \ref{th:quantum}
\begin{itemize}
    \item replaces Dirac's wave approximation (\ref{eq:Dirac}) and approximate quasi-classical expansions \cite{landau_quantum} by an exact computation, for any action with $ \ \hbar \ \Delta_{\bf M} \phi \not\approx 0$. Note that $\Delta_{\bf M} \phi$ of (\ref{eq:Delta}) can become large or even unbounded for small $M$ close to constraints or singularities, i.e., in regions where most quantum phenomena occur.
    \item uses only the $\mathbb{J}$-valued classical multipaths or actions from Theorem \ref{th:Hamilton}, which are a subset of all zig-zag paths in Feynman's path integral 
    \begin{equation}
        \psi({\bf x}_o, {\bf x}, t)= \frac{1}{Z} \int_{{\bf x}_o}^{\bf x} e^{\frac{i  }{\hbar} \int_o^t L d \theta} \ \mathcal{D} {\bf x} \label{eq:Feynman}
    \end{equation}
    where $\mathcal{D} {\bf x}$ denotes the integration over $\infty^{\infty}$ stochastically time-sliced zig-zag paths and $Z$ is the normalization factor~\cite{Feynman}.  This also applies to more recent important developments such as Duru and Kleinert's time reparameterization \cite{Duru,Kleinert2009,Pelster,Sakoda,Fujikawa}. Note that the Feynman propagator (\ref{eq:Feynman}) is formally a quantum wave solution $\psi({\bf x}_o, {\bf x}, t)$ of the Schr\"odinger equation for given ${\bf x}_o \ $, ${\bf x}$, and $t$. 
     \item extends Feynman's key result on Gaussian integrals of quadratic least actions in $\mathbb{R}^N$ \cite{Feynman} to general least actions in a constrained subset $\mathbb{G}^N \subseteq  \mathbb{R}^N$. 
\end{itemize}

Wave collapse at measurement \cite{Born1926ZurQD, Cohen-Tannoudji} is a natural consequence of our formulation. It can be derived directly from the change of the classical density $\sqrt{\rho_j}({\bf x})$ in Theorem \ref{th:Hamilton} due to a classical measurement, as we now discuss. Intuitively, a measurement transforms the classical density distribution simply into a Dirac distribution in measurement coordinates, leading to the determination of $\mathbb{J}$ classical multipaths (\ref{eq:branchpointset}) from $\{ {\bf x}_o \ \veebar \ {\bf p}_o \}$ to that measurement, and thus through (\ref{eq:Wave}) to a wave collapse in the corresponding $\psi$ .

Let us map the classical density $\sqrt{\rho_j}({\bf x})$ to $\sqrt{\rho_j}({\bf y}) =  \widehat{U}  \sqrt{\rho_j}({\bf x}) \ $, where the operator $\widehat{ {U}}$ is unitary. We use the notation $ \ \widehat{ {U}} \ a  = \int_{{\bf x} \in \mathbb{G}^N} {U}({\bf x}, {\bf y}) \ a({\bf x}) \ d x^1, ..., dx^N \ $  with ${\bf y} \in \mathbb{R}^N $, where for instance  ${U}({\bf x}, {\bf y}) = \delta({\bf x} - {\bf y})$ implies a spatial density distribution $ \sqrt{\rho_j}({\bf y}) =  \sqrt{\rho_j}({\bf x})$, whereas $ \ {U}({\bf x}, {\bf y})  = \frac{1}{\sqrt{2 \pi \hbar}} \ e^{-\frac{i {\bf y} {\bf x}}{\hbar}}$ yields the Fourier transform and thus implies a momentum ${\bf y}$ density distribution.

Consider now a classical measurement ${\bf y}_k$ of ${\bf y}$  in the density distribution $\sqrt{\rho_j}({\bf y})$. Since this classical measurement tells us the exact ${\bf y}$, which was not known before the measurement, it turns the general classical density distribution $\sqrt{\rho_j}({\bf y})$ before the measurement into the Dirac impulse $\sqrt{\Rho_k}({\bf y}) =  \widehat{ {U}}  \sqrt{\Rho_k}({\bf x}) = \delta({\bf y} - {\bf y}_k)$ after the measurement. From (\ref{eq:Wave}), this change of the classical density in turn implies that at the measurement, a general wave collapses into a wave $ \ \Psi_k({\bf y}) =  \widehat{ {U}}  \Psi_k ({\bf x}) = \delta({\bf y} - {\bf y}_k) \ \sum_{j \in \mathbb{J}} e^{\frac{i}{\hbar}\phi_j}\ $. The factor   $\sum_{j \in \mathbb{J}} e^{\frac{i}{\hbar}\phi_j}$  disappears when $\Psi_k({\bf y})$ is normalized.

As discussed in Section~\ref{col}, the Hamilton-Jacobi formulation in Theorem \ref{th:Hamilton} determines the $\mathbb{J}$  multipaths (\ref{eq:branchpointset}) to all final positions ${\bf x} \in \mathbb{G}^N$ from either an initial  ${\bf x}_o$ or an initial ${\bf p}_o \ $. This formulation is not deterministic, since at the initialization either ${\bf p}_o$ or ${\bf x}_o$ is undefined. This classical initial ambiguity is resolved when the final ${\bf y}_k$ is measured, i.e., at the measurement time the 
$\mathbb{J}$ classical multipaths (\ref{eq:branchpointset}) from $\{ {\bf x}_o \ \veebar \ {\bf p}_o \}$  to ${\bf y}_k$ are determined.

\begin{lemma}
Consider a Hermitian quantum measurement operator $\widehat{\bf Y}  = \widehat{U}^{\dagger} \ \rm{diag}({\bf y})  \ \widehat{U} \ $, based on  a measurement ${\bf y} \in \mathbb{R}^N$ and a unitary operator $\widehat{U}({\bf x}, {\bf y})$. 
Let $\sqrt{\rho_j}({\bf x})$ be the classical density distribution at measurement time $t$, with ${\bf x} \in \mathbb{G}^N$. Mapping $\sqrt{\rho_j}({\bf x})$ to $\sqrt{\rho_j}({\bf y}) =  \widehat{ {U}}  \sqrt{\rho_j}({\bf x})$, a specific measurement result ${\bf y}_k$ of ${\bf y}$  implies  that
\begin{itemize}
    \item the classical square root density distribution $\sqrt{\rho_j}({\bf y})$ before the measurement ${\bf y}$ turns into $\sqrt{\Rho_k}({\bf y}) =  \widehat{ {U}}  \sqrt{\Rho_k}({\bf x}) = \delta({\bf y} - {\bf y}_k)$ after the measurement. Since $\sqrt{\rho_j}({\bf y})$ is constrained to ${\bf y}_k$ at the measurement, $\sqrt{\Rho_k}({\bf y}) = \delta({\bf y} - {\bf y}_k)$ is a branch point of Theorem \ref{th:Hamilton}.
    \item the quantum wave $\psi({\bf y})$ before the measurement collapses into the normalized eigenwave $ \ \Psi_k({\bf y}) =  \widehat{ {U}}  \Psi_k ({\bf x}) =  \delta({\bf y} - {\bf y}_k)$.
    The wave $\Psi_k({\bf y})$ is a solution of  $ \ \rm{diag}({\bf y}) \ \Psi_k({\bf y}) = \rm{diag}({\bf y}_k) \ \Psi_k({\bf y}) \ $, which  can be written equivalently  as $ \ \widehat{\bf Y}\ \Psi_k({\bf x}) = \rm{diag}({\bf y}_k) \  \Psi_k({\bf x})$.
\end{itemize}
\noindent At measurement time, the initial condition in ${\bf x}_o$ and ${\bf p}_o$ of Theorem \ref{th:Hamilton} is completed by the measurement ${\bf y}_k \ $, which determines in Theorem \ref{th:Hamilton} the $\mathbb{J}$ classical multipaths (\ref{eq:branchpointset})  from $\{ {\bf x}_o \ \veebar \ {\bf p}_o \}$ to ${\bf y}_k \ $.
\label{lem:measurement}
\end{lemma}
In the formulation above, all eigenvalues are simple. Multiple eigenvalues can occur if $\rm{dim}\ {\bf y} < N\ $. For instance, the temporal operator $ \ {U}(t, y)  = \frac{1}{\sqrt{2 \pi \hbar}} \ e^{-\frac{i t y }{\hbar}}$ generates the temporal Fourier transform, and thus yields the density distribution of a scalar energy $y \ $  \cite{Feynman}. The classical density becomes $\sqrt{\Rho_k}(y) =  \widehat{ {U}}  \sqrt{\Rho_k}({\bf x}) = \delta(y - y_k)$ after the measurement, and the wave collapses to the eigenwave $ \ \Psi_k(y) =  \widehat{ {U}}  \Psi_k ({\bf x}) =  \delta(y - y_k)$. In this case, multiple eigenwaves $\Psi_k ({\bf x})$ correspond to the same eigenvalue $y$.

Note that Theorem \ref{th:quantum} and Lemma \ref{lem:measurement} directly derive  fundamental postulates of quantum mechanics~\cite{Cohen-Tannoudji} from the classical physics in Theorem \ref{th:Hamilton}. Namely, (i) the quantum system state is represented as a wave function $\psi_j({\bf x}, t)$, now derived from the classical action $\phi_j({\bf x}, t)$, (ii) the wave function solves the Schr\"odinger equation (both from Lemma~\ref{lem:equivalence} and Theorem \ref{th:quantum}), (iii) which implies wave collapse at measurements (from Lemma \ref{lem:measurement}).

Born's measurement rule \cite{Born1926ZurQD, Cohen-Tannoudji} remains a postulate. It states that the probability to measure the position ${\bf x}$ is the (now classically derived) scalar quantum density $\varrho ({\bf x}, t)$ of (\ref{eq:Prob}) in Theorem \ref{th:quantum} , or in the matrix case  the diagonal elements of $\varrho ({\bf x}, t)$. 
It implies that the probability of measuring ${\bf y}_k$ in Lemma \ref{lem:measurement} from the wave $ \ \psi({\bf y}) = \sum_{k \in \mathbb{N}} c_k \Psi_k({\bf y}) \ $ is $ \ \varrho_{{\bf y}_k} = \sum\limits_{k \ with \ same \ {\bf y}_k} c_k^{\ast} c_k \ $.

Also note that Lemma \ref{lem:measurement} suggests the possibility of a classically-based interpretation of wave collapse at a quantum measurement, as an alternative to the Copenhagen interpretation \cite{bohr1928quantenpostulat, Heisenberg1925, Schrodinger1926, heisenberg1958physics}.
The Copenhagen interpretation suggests that this decision is taken at the measurement time $t$. The classically-based interpretation would be that these decisions are taken before the measurement, at the initial condition $t=0$ and at the branch points along the $\mathbb{J}$  multipath. Which interpretation best describes the nature of physical reality remains an open question since both interpretations lead to the same experimental results.
However, Lemma \ref{lem:measurement} is fully derived from and consistent with the  Hamilton-Jacobi and Euler formulations of Theorem \ref{th:Hamilton}.

Finally, note that for periodic waves a Bohr-like quantization rule \cite{Bohr1913} can be derived.
\begin{lemma}
\ \ Given periodic waves $ \ \psi_j = \sqrt{\rho}({\bf x}, t) \ e^{\frac{i }{\hbar} (\phi({\bf x}, t) \ + \ k \  \varphi(\omega))} \ $ in  Theorem \ref{th:quantum}, with 
$ \ \mathbb{J} = \{ \omega \in \mathbb{R} \ \times \  k \in \mathbb{N} \} \ $ and $\varphi$ an arbitrary continuous function, the continuous parameter $\omega$ in the cumulated wave  $\psi$ is quantized as $ \ \frac{\varphi(\omega)}{\hbar} = 2 \pi k$.
\label{lem:Bohr}
\end{lemma}

 \noindent {\bf Proof} \ \ The cumulated wave is derived from the geometric series
\begin{equation}
  \lim_{K \rightarrow +\infty}  \frac{1}{K} \sum_{\kappa = 0 }^K e^{ i \kappa \frac{\varphi}{\hbar} }  =  \lim_{K \rightarrow +\infty} \ \ \frac{1}{K}\ \frac{1 }{1 - e^{ i K \frac{\varphi}{\hbar}  } } =   \left\{ 
\begin{array}{ll}
1 & \text{\  for any \  } \frac{\varphi}{\hbar} = 2 \pi k  \\
0 & \text{\ for any \  } \frac{\varphi}{\hbar} \ne 2 \pi k  
\end{array}
\right.\ \  k \in \mathbb{N} \nonumber
\end{equation}
All periodic actions are included in the summation, but the above implies that only $2 \pi k$-periodic actions remain. This is the source of quantization. $ \hfill \square$


\subsection*{Examples} \label{examples}

We illustrate on examples how the wave function can be systematically constructed based only on classical action and classical density. Each example first finds the multi-valued action solution of the classical Hamilton-Jacobi p.d.e. of Theorem \ref{th:Hamilton}. Position and time coordinates are chosen to simplify the derivation. Next, the classical densities are computed for each action branch. Finally, the quantum wave is constructed from Theorem \ref{th:quantum}. Basic computational tools are introduced as needed before each example. 

\begin{figure}
\begin{subfigure}[b]{0.51 \textwidth}
   \includegraphics[scale=0.51]{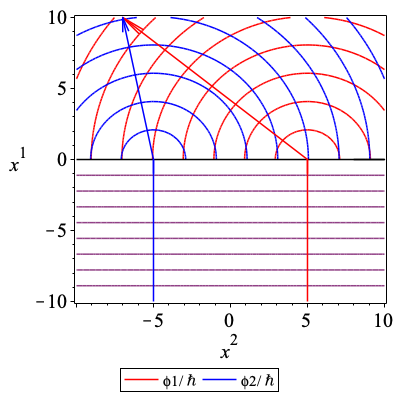}
   \caption{Actions $\phi_1$ and $\phi_2$ in the two-slit experiment} \hfill
    \label{fig:Twoslit}
\end{subfigure}
\hspace{0.2in}
\begin{subfigure}[b]{0.49 \textwidth}
    \includegraphics[scale=0.5]{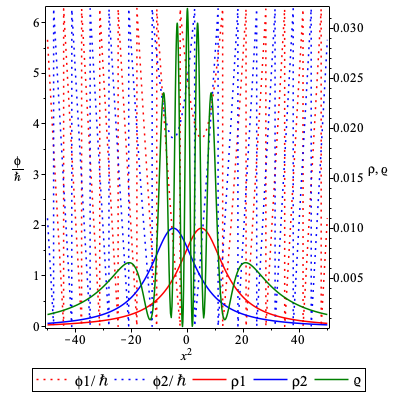}
    \caption{Classical actions and densities on the screen, and associated quantum probability density}
    \label{fig:Twoslitprob}
\end{subfigure}
\caption{Classical particle paths and quantum wave in the double-slit experiment} \label{fig:TwoslitAll}
\end{figure}

\Example{Double slit experiment}{\ \ Consider the classical Hamiltonian-Jacobi p.d.e. (\ref{eq:HJ}) on the two-connected manifold $\mathbb{G}^3$ in Figure \ref{fig:Twoslit}
\begin{equation}
- {\frac {\partial \phi_j}{\partial t}} = H  =  \frac{1}{2 M} \nabla \phi_j^T \nabla \phi_j \ \ \ \ \ \ \ \ \ \ \ \ \ \ \ 
{\bf x}= (x^1, x^2, x^3)^T \in \mathbb{G}^3 = \mathbb{R}^3 \setminus \mathbb{W}^3 \nonumber  
\end{equation}
with constant mass $M$, wall with two holes $ \ \mathbb{W}^3 = \{ x^1 = 0  \}  \setminus  \mathbb{H}^3$, where  $\mathbb{H}^3 = \{ {\bf x}_1, {\bf x}_2 \}$ with ${\bf x}_{1, 2} = (0, \pm 5, 0)^T $, and initial momentum $p_o$ of the particle (for instance, an electron). 
Letting $r_j = \sqrt{ ({\bf x} - {\bf x}_j)^T({\bf x} - {\bf x}_j)}$ after the wall, the $\mathbb{B} = \{1, 2 \}$ - valued least action of Theorem \ref{th:Hamilton} is
\begin{equation}
{\phi}_j  =
\left\{ 
\begin{array}{ll}
p_o \ x^1 - E t & \text{for \ }  x^1 < 0 \\
p_o \ r_j - E t & \text{for \  }  x^1  \ge 0  
\end{array}
\right. \ \ \ \ \ \ \ \ \ \ \ \ \ {\rm where} \ \ E = \frac{p_o^2}{2 M} \nonumber
\end{equation}
They are shown in Figure \ref{fig:Twoslit}. Since $H = E \ $ is constant and there is no potential term, in this example the action simply corresponds to the geometric distance. Using in (\ref{eq:Delta}) the spherical Laplacian $ \ \Delta_{\bf M} f(r) = \frac{1}{M r^2} \frac{\partial}{\partial r} (r^2 \frac{\partial f}{\partial r}) =  \frac{1}{M r} \frac{\partial^2}{\partial r^2} (r f) \ $, the classical density (\ref{eq:density}) is
\begin{equation}
\Delta_M \phi_j = 
\left\{ 
\begin{array}{ll}
0   & \text{  for  } x^1 < 0 \\
\frac{2}{r_j} \frac{p_o}{M} =  \frac{\dot{r}_j}{r_j}& \text{ for  } x^1 \ge 0  
\end{array}
\right.  \ \ \ \ \implies \ \ \ \ 
\rho_j = 
\left\{ 
\begin{array}{ll}
1 & \text{  for  } x^1 < 0 \\
\frac{1}{r_j^2} 
& \text{ for  } x^1 \ge 0  
\end{array}
\right. \label{eq:densitytwoslit}
\end{equation}
The least action branches are illustrated in Figure \ref{fig:Twoslit}. Both slits are branch points with fully elastic collision forces in (\ref{eq:dotp}) and an infinite classical density. The classical non-Lipschitz constraint forces (\ref{eq:dotp}) in the two slits lead to an infinity of radial paths (\ref{eq:dotq}) from each slit, which connect every measurement pixel $x^2$ on the screen at $x^1 = 10$ with two least action paths (\ref{eq:dotq}). Thus, the statistical density distribution $\rho({\bf x}(t))$ behind the slits is just the evolution of the Dirac densities in the slits along the density path integrals (\ref{eq:density}) after the slits. 

Theorem \ref{th:quantum} in turn yields the (un-normalized) wave function (\ref{eq:Wave}), 
\begin{eqnarray}
{\psi} &=&  \sum_{j \in \mathbb{B}}  \sqrt{\rho_j} \ \ e^{\frac{i }{\hbar} \phi_j}  =  e^{-\frac{i}{\hbar} E t}
\left\{ 
\begin{array}{ll}
 e^{\frac{i}{\hbar} p_o x^1 } & \ \text{for \ } x^1 < 0 \\
  \frac{1}{r_1}\  e^{\frac{i}{\hbar} p_o r_1 } + \frac{1}{r_2}\ e^{\frac{i}{\hbar} p_o r_2 }  & \ \text{for  \ } x^1 \ge 0  
\end{array}
\right. \ \ \ \ \ \  \label{eq:Fraunhofer}
\end{eqnarray}  
One can directly confirm that each term fulfills the Schr\"odinger equation (\ref{eq:Schroed}). When specialized to the far field, this result matches the well-known two-slit Fraunhofer wave function \cite{Fraunhofer1823}. The wave collapse in both slits can be interpreted according to Theorem \ref{th:quantum} as the transition in Figure \ref{fig:Twoslit} from the flat least action branch before the wall to the two conic least action branches after the wall. 

With $ \ r_j = \sqrt{(x^1)^2 + (x^2 \pm 5)^2} \ $, the classical actions $ \ \frac{\phi_j}{\hbar} = \frac{p_o \ r_j}{\hbar} \ [{\rm mod}\ 2 \pi]$ (where the $2 \pi$ periodic part is removed), densities $\rho_j= r_j^{-2}$ and the resulting near field probability density $\varrho  = \psi \psi^{\dagger}$ on the screen at $x^1 = 10$ are plotted in Figure \ref{fig:Twoslitprob} with $\frac{p_o}{\hbar}=2$. The quantum density $\varrho$ is a phase weighted combination of the two classical densities $\rho_j$, with the two phases derived from the actions. On the left and right sides of the Figure the two phases have a nearly constant offset, whereas in the center they significantly change which leads to the wave oscillation. Feynman's zig-zag path integrals (\ref{eq:Feynman}), which were originally motivated by this example, can now be reduced to just two paths.

The Hamilton-Jacobi formulation of Theorem \ref{th:Hamilton} cannot deterministically predict from the slits where the particle hits the screen. Indeed, in the slits only the position is given, but not the momentum, which is affected by the non-Lipschitz constraint force. However, the path from the slit to the screen is a determined straight line following (\ref{eq:dotq}) of Theorem \ref{th:Hamilton}. Lemma \ref{lem:measurement} thus allows a classically-based interpretation that the decision where the particle hits the screen is already taken in the slits from the non-Lipschitz constraint force (\ref{eq:dotq}) of Theorem \ref{th:Hamilton}, i.e.,  before the final position is measured on the screen.

Also, measuring the particle position in slit 2 with a photon, for instance, would change the branch index from $\ j=1,2 \ $ to $\ j=2\ $, leading to a single classical action cone behind the wall. Note that in principle this example extends to slits or holes of finite width in $\mathbb{H}^3$, extending the summation in (\ref{eq:Fraunhofer}) to all elements of $\mathbb{H}^3$. A particle described, e.g., with Laguerre polynomials~\cite{zwiebach2022} will always collide with the edge of a slit, with the wave function (\ref{eq:Fraunhofer}) extending accordingly to the sum over all points of both slits.}{Twoslit}

\Example{Aharonov–Bohm effect}{ \ \ Another example is the Aharonov–Bohm effect~\cite{Aharonov1959}, where a charged particle is affected by the magnetic potential ${\bf A}$ in the absence of any actual magnetic field,  ${\bf B} = \mathrm{rot} \ {\bf A} = {\bf 0}$. The presence of ${\bf A}$ in the Hamiltonian does affect the classical action in the Hamilton-Jacobi p.d.e. (\ref{eq:HJ}) and thus the quantum wave constructed in Theorem \ref{th:quantum}.

Consider for instance the two-slit set-up of \Ex{Twoslit}, with the same  two-connected manifold $\mathbb{G}^2$ as in Figure \ref{fig:Twoslit}, but assume now that behind the wall ($x_1 > 0$) there is a magnetic potential ${\bf A}({\bf x})$, with gauge $\nabla_M \cdot {\bf A} = 0$. The classical Hamiltonian-Jacobi p.d.e. (\ref{eq:HJ}) is
\begin{equation}
    \ -  \frac{\partial \phi_j}{\partial t} =  H  = \frac{1}{2 M} \left( \nabla \phi_j - Q \ {\bf A} \right)^T \left( \nabla \phi_j - Q \ {\bf A} \right)  \ \ \ \ \ \ \ \ \ \ \  \ \ \ \ \ \   {\bf x}= (x^1, x^2, x^3)^T \in \mathbb{G}^3 = \mathbb{R}^3 \setminus \mathbb{W}^3   \nonumber \\
\end{equation}
with mass $M$, particle charge $Q$, wall with two holes $ \ \mathbb{W}^3 = \{ x^1 = 0  \}  \setminus  \mathbb{H}^3$, where  $\mathbb{H}^3 = \{ {\bf x}_1, {\bf x}_2 \}$ with $ \ {\bf x}_{1, 2} = (0, \pm 5, 0)^T $, and initial momentum $p_o \ $. The $\mathbb{B} = \{1, 2 \}$ - valued least action of \Ex{Twoslit} becomes  
\begin{equation}
{\phi}_j  = 
\left\{ 
\begin{array}{ll}
\ p_o \ x^1 - E t & \ \ \ \ \text{for \ }  x^1 < 0 \\
\ p_o \ r_j - E t \ +  Q \int_{{\bf x}_j}^{\bf x} {\bf A} \ d {\bf x}& \ \ \ \ \text{for \  }  x^1  \ge 0  
\end{array}
\right. \ \ \ \ \ \ \ \ \ \ \ \ \ {\rm where} \ \ E = \frac{p_o^2}{2 M} \nonumber
\end{equation}
Note that the integral above is path-independent since ${\bf A}$ is a gradient field (by hypothesis, $\mathrm{rot} \ {\bf A} = {\bf 0}$). The vanishing magnetic field ${\bf B}$ implies that the classical particle path and hence the classical density in (\ref{eq:densitytwoslit}) are identical to those of the two-slit experiment in Figure \ref{fig:TwoslitAll}, Theorem \ref{th:quantum} now yields the (un-normalized) wave function (\ref{eq:Wave}), 
\begin{eqnarray}
{\psi} &=&  \sum_{j \in \mathbb{B}}  \sqrt{\rho_j} \ \ e^{\frac{i }{\hbar} \phi_j}  =  e^{-\frac{i}{\hbar} E t}
\left\{ 
\begin{array}{ll}
 e^{\frac{i}{\hbar} p_o x^1 } & \ \ \text{for \ } x^1 < 0 \\
  \frac{1}{r_1}\  e^{\frac{i}{\hbar} ( p_o r_1 + Q \int_{{\bf x}_1}^{\bf x}  {\bf A} d {\bf x} )} + \frac{1}{r_2}\ e^{\frac{i}{\hbar} ( p_o r_2 + Q \int_{{\bf x}_2}^{\bf x} {\bf A} d {\bf x}) }  & \ \ \text{for  \ } x^1 \ge 0  
\end{array}
\right. \nonumber
\end{eqnarray} 
Hence Theorem \ref{th:quantum} provides a classically-based explanation of the Aharonov–Bohm phase shift \cite{Aharonov1959}, which is added in the Fraunhofer wave (\ref{eq:Fraunhofer}).}{Aharonov}

\begin{figure}
    \centering
    \includegraphics[scale=0.5]{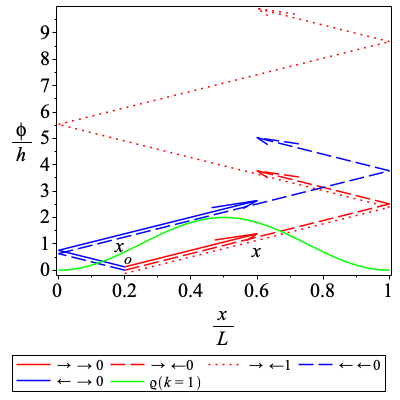}
    \caption{Multipaths from $\frac{x_o}{L}=0,2$ to $\frac{x}{L}=0,6$ in a box, and the resulting wave.} 
    \label{fig:Particle}
\end{figure}

\Example{Particle in a  box}{\ \ Consider the Hamilton-Jacobi p.d.e. (\ref{eq:HJ}) of a particle in a box of width $L$ in Figure \ref{fig:Particle}
$$
- \frac{\partial \phi}{\partial t} = H =  \frac{1}{2 M} \nabla \phi^2 \ \ \ \ \ \ \ \ \ \ \ 0 \le \frac{x}{L} \le 1
$$ 
with position $x$, initial position $x_o$, constant mass $M$, and unknown momentum $p$. The $\mathbb{B} = \{ \{ \rightarrow, \leftarrow \}  \ \times \      \{ \rightarrow, \leftarrow \}  \ \times \ p \in \mathbb{R}_{+} \ \times \  k \in \mathbb{N} \}$ - valued action of Theorem \ref{th:Hamilton} 
\begin{equation}
{\phi}_{j} \ = \ 
2 k \ L  p - E t + \left\{ 
\begin{array}{ll}
p  \left( x - x_o \right)  & \text{for \  } j = \ \rightarrow \rightarrow k \\
p  \left( 2 L - (x + x_o) \right) & \text{for \  } j = \ \rightarrow \leftarrow k \\
p  \left( 2 L - (x - x_o) \right)  & \text{for \  } j = \ \leftarrow \leftarrow k \\
p  \left( x + x_o \right)  & \text{for \  } j = \ \leftarrow \rightarrow k  
\end{array}
\right. \ \ \ \ \ \ \ \ \ E = \frac{p^2}{2 M} \nonumber
\end{equation}
is illustrated in Figure \ref{fig:Particle}. The paths are augmented with $2 k \ L$ periodic path elements since each wall reflection leads to an additional action branch. From Lemma \ref{lem:Bohr}, this implies the quantization $2 \frac{L p}{\hbar} \ = \ 2 \pi k,\ k  \in \mathbb{N}$, yielding $p =  \pi k \frac{\hbar}{L}$ so that the number of solutions reduces to $\mathbb{B} =  \{ \{ \rightarrow, \leftarrow \}  \ \times \  \{ \rightarrow, \leftarrow \}  \ \times \   k \in \mathbb{N}  \}$. Using $\Delta_M \phi_{k} = 0$ outside the edges, the classical density (\ref{eq:density}) $\sqrt{\rho_{k}} = \sqrt{\frac{2}{L}} \frac{1}{2 i }$ is constant.
Thus, letting $E_k = \frac{\hbar^2 \pi^2 k^2}{2M L^2}\ $, the normalized wave function (\ref{eq:Wave}) of Theorem \ref{th:quantum} is
\begin{eqnarray}
\psi \  = \ \sum_{k \in \mathbb{N}}   \sqrt{\rho_{k}} \ \sum_{\{ \rightarrow , \leftarrow \}  \ \times \   \{ \rightarrow , \leftarrow \}  }  
\ e^{\frac{i }{\hbar} \phi_{j}} \ = \ \sqrt{\frac{2}{L}} \sum_{k \in \mathbb{N}^\star}   e^{- \frac{i}{\hbar} E_k t } 
\sin {\frac{\pi k x_o}{L}} \ \sin { \frac{\pi k x}{L} }  \nonumber
\end{eqnarray}
Figure \ref{fig:Particle} illustrates for $k= 1$ the 4 classical "billiard" paths. In addition, one periodic extended path $\rightarrow  \rightarrow  1$ is shown. Measuring the particle energy $E_k$ corresponds to selecting a branch $k$ in Theorem \ref{th:Hamilton}. From Lemma \ref{lem:measurement}, this is associated to a wave collapse. The resulting probability density $\varrho = \psi \psi^{\dagger}$ is also shown for $k=1$. The novelty is that this well-known result \cite{Schrodinger1926, zwiebach2022} is derived just from the constrained, multi-valued action of Theorem \ref{th:Hamilton}.}{box}

\Example{Tunneling}{\ \ Consider the barrier problem with the Hamilton-Jacobi p.d.e. (\ref{eq:HJ}),
\begin{equation}
   - \frac{\partial \phi}{\partial t} = H =  \frac{1}{2 M} \nabla \phi^2  + V \ \ \ \ \ \ \ \ {\rm where}\ \  V(x) = \left\{ 
\begin{array}{ll}
0 & \ \text{for \  } x < 0 \\
V_{\infty} > 0 & \ \text{for \  } x \ge 0 
\end{array}
\right. \ \ \ \ \ \ \ \ x \in \mathbb{R}  \label{eq:tunneling}
\end{equation}
where particles with constant mass $M$ are shot with an initial momentum $p_o$ and density $\rho_o$ toward the potential barrier $V(x)$ at a position $x$. Besides the initial path, for $x < 0$ there is a reflected path $R$, and for $x \ge 0$ the possibility of a transmitted path $T$. The $\mathbb{B} = \{ \rightarrow, \leftarrow \} $ - valued action of Theorem \ref{th:Hamilton} is

\begin{equation}
{\phi}_{j} \ = \ 
\left\{ 
\begin{array}{ll}
- \frac{p_o^2}{2 M} \ t + p_o \ x & \text{for \  } x < 0, \ j = \rightarrow  \\
- \frac{p_o^2}{2 M} \ t - p_o \ x & \text{for \  } x < 0, \ j = \leftarrow  \\
- \frac{p_T^2}{2 M} \ t + p_T \ x & \text{for \  } x \ge 0, \ j = \leftarrow, \rightarrow  
\end{array}
\right.  \nonumber
\end{equation}
where $ \ p_T = \sqrt{ p_o^2 - 2 M \ V_{\infty}}$ is real for $ \ \frac{p_o^2}{2 M} \ge V_{\infty} \ $ and imaginary for $ \ \frac{p_o^2}{2 M} < V_{\infty} \ $, i.e., a complex action exists in both cases. The action is real except for $x > 0$ and  $\frac{p_o^2}{2 M} < V_{\infty} $ where it is imaginary. 

The position dynamics (\ref{eq:dotp}, \ref{eq:dotq}) of (\ref{eq:tunneling}) is linear for $x > 0$. Thus, superposition of the two imaginary path solutions $ \ M \dot{x}_{1, 2} = \pm p_T \ $ yields  the real position solution $x = x_1 + x_2 \ $, with real velocity  $\dot{x} = 0$. The conjugate imaginary actions and momenta are well defined solutions of the Hamilton-Jacobi equation (\ref{eq:HJ}), and correspond by superposition to a real position with zero velocity.

Since $\Delta_M \phi_j = 0$, the classical density (\ref{eq:density}) is constant for $x \ne 0$. At $x=0$, the classical continuity equation (\ref{eq:continuity}) and force equilibrium imply
$$ 
\frac{p_o}{M}   \ \rho_o  =  \frac{p_o}{M} \ \rho_T + \frac{p_T}{M} \ \rho_R, \ \ \ \left( \rho_T + 2 \rho_R \right) \left(\frac{p_o}{M}\right)^2 = \left( \rho_o + \rho_R \right) \left(\frac{p_o}{M}\right)^2 = 
\left( \left(\frac{p_o}{M}\right)^2 + \left(\frac{p_T}{M}\right)^2 \right) \rho_T
$$
The normalized wave function (\ref{eq:Wave}) of Theorem \ref{th:quantum} or Lemma \ref{lem:equivalence} for the complex case is hence
\begin{eqnarray}
\psi \  = \sum_{j \in \mathbb{B}}    \sqrt{\rho_{j}} \ e^{\frac{i }{\hbar} \phi_{j}} \ = \  \left\{ 
\begin{array}{ll}
\left(\sqrt{\rho_o} \ e^{\frac{i }{\hbar} p_o \ x} + \sqrt{\rho_R} \ e^{- \frac{i }{\hbar} p_o \ x} \right) \ e^{- \frac{i }{\hbar} \frac{p_o^2}{2 M} \ t}  & \ \text{for \  } x \le 0 \\
\sqrt{\rho_T} \ e^{\frac{i }{\hbar} p_T \ x - \frac{i }{\hbar} \frac{p_T^2}{2 M} \ t} & \ \text{for \  } x > 0 
\end{array}
\right.\nonumber
\end{eqnarray}
Again, the novelty is that this well-known wave function \cite{Hund1927tunneling, Devoret1985, zwiebach2022} is derived just from the classical multi-valued action and density of Theorem \ref{th:Hamilton}, covering the "classically prohibited" area.}{barrier}

The next examples use Hermite polynomials $H_k(z) = \left( 2 z - \frac{d}{dz} \right)^k \cdot 1,  k \ge 0$ which yield basis functions orthonormal with respect to the measure $e^{-z^2}$  \cite{zwiebach2022}
\begin{equation}
\Psi_{k_n}(x^n) =  \sqrt[4]{\frac{M  \omega}{\pi \hbar}} \frac{1}{\sqrt{2^{k_n} k_n!}} \
H_{k_n} (z^n) e^{-\frac{1}{2} (z^n)^2}\ \ \ \ \ \ \ \ {\rm with} \ \ \ \ \ z^n = x^n \sqrt{\frac{M  \omega}{\hbar}} \ \ \ \ \ \ \label{eq:HermitePsi}
\end{equation}

\Example{Harmonic oscillator}{\ \ Consider the Hamilton-Jacobi p.d.e. (\ref{eq:HJ}) of Theorem \ref{th:Hamilton}
\begin{eqnarray}
   - \frac{\partial \phi}{\partial t} = H \ = \frac{1}{2} \left( \frac{\nabla \phi^T \nabla \phi }{M} \ + \ M  \omega^2  \ {\bf x}^T {\bf x} \right) \ \ \ \ \ \ \ \ \ \ {\bf x} = (x^1, ..., x^N) \in \mathbb{R}^N, \  t \ge 0 \nonumber 
\end{eqnarray}
with constant angular frequency $\omega$ and mass $M$, Cartesian position ${\bf x}$ and given initial position ${\bf x}_o$. The single-valued real least action of Theorem \ref{th:Hamilton} is
\begin{equation}
 \phi = \frac{M  \omega}{2} \left( \cot{\omega t} \ ({\bf x}^T {\bf x} + {\bf x}_o^T {\bf x}_o )  - \frac{2}{\sin{\omega t}}  \ {\bf x}^T {\bf x}_o   \right) \label{eq:harmaction}
\end{equation}
which solves the Hamilton-Jacobi p.d.e. \cite{Feynman}, with $ \ \phi({\bf x} = {\bf x}_o\ , t= 2 k \pi) = 0 \ $ for $k \in \mathbb{Z} \ $ from L'Hôpital's rule. Using $\ \Delta_M \phi = N \omega \cot{\omega t} \ $, the classical density (\ref{eq:density}) can be computed as $\sqrt{\rho}({\bf x}, t) = \sqrt{\frac{M  \omega}{2 i \pi  \hbar \sin{\omega t } }}^N$.  

The normalized wave function (\ref{eq:Wave}) of Theorem \ref{th:quantum} is
\begin{eqnarray}
\psi &=&  \sqrt{\rho }   e^{\frac{i }{\hbar} \phi  } 
 = \sum_{k \in \mathbb{N}}^{\forall k_1+ ...+ k_N = k}    e^{-\frac{i}{\hbar} E_k  t} \ \prod_{n=1}^N   \Psi_{k_n}(x^n) \Psi_{k_n}(x_o^n) \ \ \ \ \ \ \ \ \ \label{eq:eigenstatexxo} 
\end{eqnarray}
where we used a Taylor series expansion, as detailed in \cite{Feynman}.
This leads to the eigenvalues $E_k = \hbar \omega (k + \frac{N}{2})$ and wave eigenfunctions (\ref{eq:HermitePsi}).

The computation above exploits many elements of the Feynman path integral~\cite{Feynman}, but there is no intrinsic process noise added to the classical path. Feynman~\cite{Feynman} showed that no process noise is needed in the particular case of a single-valued quadratic action in ${\bf x} \in \mathbb{R}^N$, but did not extend this result to a non-quadratic multi-valued action in a constrained manifold, as is the case in all other examples of this paper. }{Harmonic}
The next example uses quaternion coordinates $\pm{\bf q}({\bf x})= \pm (q^1, ..., q^4)$, which relate to Cartesian coordinates ${\bf x} = (x^1, x^2, x^3) $  \cite{Goldstein} as
\begin{eqnarray}
x^1 &=& \ 2 q^1 q^3 + 2 q^2 q^4 \nonumber \\
x^2 &=& \ -2 q^1 q^2 + 2 q^3 q^4  \label{eq:quaternion} \\
x^3 &=& \  (q^1)^2 - (q^2)^2 - (q^3)^2 + (q^4)^2  \nonumber 
\end{eqnarray}
For the $2$-dimensional case, i.e., $q_3, q_4 = 0$, the above is a complex square root with $q^1 = \pm \sqrt{\frac{1}{2}(\sqrt{(x^2)^2 + (x^3)^2} + x^3)}$, $ q^2 = \pm sign(x^2) \sqrt{\frac{1}{2}(\sqrt{(x^2)^2 + (x^3)^2} - x^3)}$. The kinetic energy can be transformed from Cartesian to quaternion coordinates as $M \ \dot{\bf x}^T \dot{\bf x} = M \ \dot{\bf q}^T \frac{\partial {\bf x}}{\partial {\bf q}}^T \frac{\partial {\bf x}}{\partial {\bf q}} \dot{\bf q} = 4 M \ {\bf q}^T {\bf q}  \ \ \dot{\bf q}^T  \dot{\bf q}$. Using these coordinates, we now show how the action and wave of an electron around a proton Coulomb field derive from the harmonic oscillator, yielding the basic model of a hydrogen atom.

\begin{figure}
\begin{subfigure}[t]{0.49\textwidth}
\centering
    \includegraphics[width=5cm, height=4.2cm]{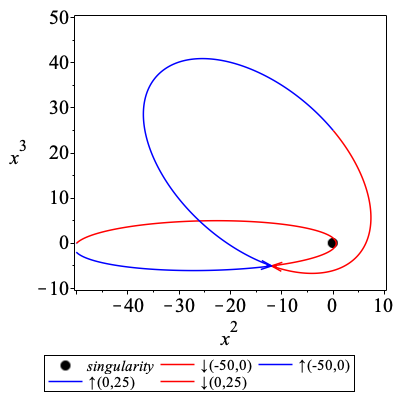}
    \caption{Quantized Kepler orbit pairs} 
    \label{fig:Coulomb}
\end{subfigure}
\hspace{0.02cm}
\begin{subfigure}[t]{0.49 \textwidth}
\centering
    \includegraphics[width=5cm, height=4.2cm]{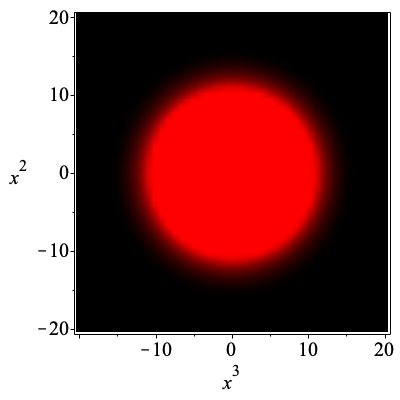}
    \caption{Orbit 1S: $c_{k_1=k_2=0}=1$}
    \label{fig:CoulombProb0}
\end{subfigure} \\
\begin{subfigure}[t]{0.49\textwidth}
\centering
    \includegraphics[width=5cm, height=4.2cm]{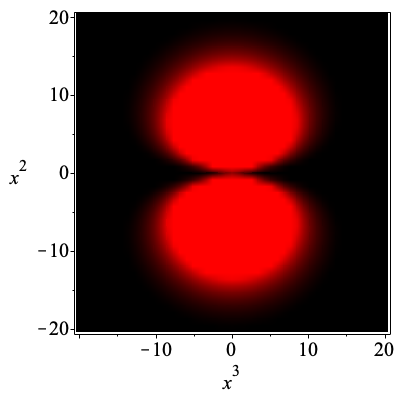}
    \caption{Orbit 2P: $c_{k_1=k_2=1}=1$}
    \label{fig:CoulombProb}
\end{subfigure}
\hspace{0.02cm}
\begin{subfigure}[t]{0.49 \textwidth}
\centering
    \includegraphics[width=5cm, height=4.2cm]{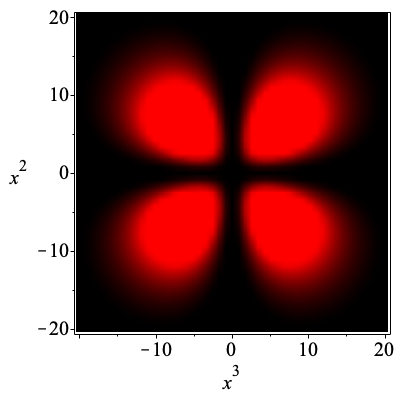}
    \caption{Orbit 3D: $c_{k_1=1, k_2=3}=1, c_{k_1=3, k_2=1}=-1$}
    \label{fig:CoulombProb2}
\end{subfigure}
\caption{Quantized Kepler orbits and hydrogen orbitals}
\end{figure}

\Example{Coulomb or gravity potential}{\ \  Consider a particle in Figure \ref{fig:Coulomb} with the Hamilton-Jacobi p.d.e. (\ref{eq:HJ}) of Theorem \ref{th:Hamilton} 
\begin{eqnarray}
-  \frac{\partial \phi}{\partial t} &=&  H   = \frac{1}{2 M} \frac{\partial \phi}{\partial {\bf x}}^T \frac{\partial \phi}{\partial {\bf x}} - \frac{G}{r}  \ \ \ \ \ \ \ \ \ \ \  \ \ \ \ \ \  \ \  {\bf x} = (x^1, ..., x^3) \in \mathbb{R}^3, \  t \ge 0      \nonumber \\
      &=& \frac{1}{r} \left( \frac{1}{2 \cdot 4 M} \frac{\partial \phi}{\partial {\bf q}}^T \frac{\partial \phi}{\partial {\bf q}}   - G \right) \ \ \ \ \ \ \ \ \ \ \  \ \ \  {\bf q} = (q^1, ..., q^4) \in \mathbb{R}^4, \  t \ge 0     \nonumber \\
  -  \frac{\partial \phi}{\partial t'}  &=&  \frac{1}{2 \cdot  4 M} \frac{\partial \phi}{\partial {\bf q}}^T \frac{\partial \phi}{\partial {\bf q}}- G\ \ \ \ \ \ \ \ \ \ \  \ \ \ \ \ \ \ \ \ \ \ \ \   t' =\int_o^t \frac{dt}{r} \ \ \ \ \ \ \ \ \ \ t =\int_o^{t'} r \ dt' \ \   \label{eq:HJBGravity}
\end{eqnarray} 
with the two-valued quaternion $\pm {\bf q}({\bf x})$ of (\ref{eq:quaternion}) initialized at $\pm {\bf q}_o\ $, radius $r = {\bf q}^T {\bf q} = \sqrt{{\bf x}^T {\bf x}}\ $, and constant particle mass $M$. The constant
$G$ corresponds to the Newtonian constant of gravitation scaled with the constant mass of the particle and the mass of the singularity, or equivalently to the Coulomb constant scaled with the constant charge of the particle and the charge of the singularity.

The singularity at the origin is a branch point with a $\mathbb{J} = \{ {\bf q}_o \in \mathbb{R}^4 \} \ \times \ \{ \{ \uparrow, \downarrow \}   \ \times \  \omega \in  \mathbb{R}_{+}^{\star}  \ \times \  k \in \mathbb{N}   \}    $ - valued real least action \cite{Duru, Kleinert2009} of Theorem \ref{th:Hamilton},
\begin{equation}
\phi_j = \frac{4 M \omega}{2}  \left( ( {\bf q}^T {\bf q} + {\bf q}_o^T {\bf q}_o ) \cot{(\omega t')} \mp  \frac{2}{\sin{(\omega t')}}  {\bf q}^T {\bf q}_o \right) + \frac{4M}{2}\omega^2t +  G \left( t' + \frac{2 \pi k}{\omega} \right)   \nonumber
\end{equation}
which solves (\ref{eq:HJBGravity}) with $ \ \phi_j({\bf q} = {\bf q}_o\ , t= 2 k \pi) = 0 \ $ for $k \in \mathbb{Z} \ $ from L'Hôpital's rule, and the turn rate $\omega \in \mathbb{R}_{+}^{\star}$. The two-valued quaternions $\pm {\bf q}({\bf x})$ correspond in Figure \ref{fig:Coulomb} to the right-turning $\downarrow$ and left-turning $\uparrow$ Kepler orbits \cite{Kepler}. The orbits are augmented with $G \frac{2 \pi k}{\omega} $ periodic orbit elements, since each additional full rotation  $2 \pi$  leads to an additional action branch. From Lemma \ref{lem:Bohr}, $\omega$ is quantized as $\ \frac{G}{\hbar} \frac{2 \pi}{\omega} = \ 2 \pi k, \ k  \in \mathbb{N}^\star$, yielding $\omega = \frac{G}{\hbar k}\ $, so that the number of action solutions reduces to $\mathbb{J} = \{ {\bf q}_o \in \mathbb{R}^4 \} \ \times \ \{ \{ \uparrow, \downarrow \} \ \times k \ \in \mathbb{N}^\star  \}$. 

For a given ${\bf q}_o \ $,  the classical density (\ref{eq:density}) can be computed from $\Delta_M \phi_j = 4 \omega \cot{\omega t'}$ as $\sqrt{\rho_j} =  \sqrt{\frac{M  \omega}{2 i \pi  \hbar \sin{\omega t' } }}^{\ 4}$. Using (\ref{eq:eigenstatexxo}) from \Ex{Harmonic}, the normalized wave function (\ref{eq:Wave}) of Theorem \ref{th:quantum} is
\begin{eqnarray}
\psi ({\bf q}, {\bf q}_o, t) &=&  \sum_{j \in \mathbb{J}}  \sqrt{\rho_j } \ \ e^{\frac{i}{\hbar} \phi_j  } = \sum_{k' \in \mathbb{N}}^{\forall k_1+ ...+ k_4 = k'}   c_{k_1 .. k_4} \ e^{\frac{i}{\hbar} (G  - E_{k'} ) t' + \frac{i}{\hbar} E_k t} \prod_{n=1}^4 \Psi_{k_n}(q^n)  \Psi_{k_n}(q^n_o)  \nonumber \\
&=&   \sum_{k \in \mathbb{N}^\star}^{\forall k_1+ ...+ k_4 = 2k-2} c_{k_1 .. k_4} \ e^{\frac{i}{\hbar} E_k t} \prod_{n=1}^4 \Psi_{k_n}(q^n) \Psi_{k_n}(q^n_o)   \label{eq:qwave} 
\end{eqnarray} 
where we used $E_{k'}  =  \hbar \omega (k' + \frac{4}{2}) = 2 \hbar \omega k = G \implies E_k = 2 M  \omega^2 = \frac{M}{2} \left( \frac{G}{\hbar k}\right)^2$ to remove the dependence on $t'$. Note that $ \ k' +2 = 2k \ $ is even, due to the symmetry $H_{k_n} (-q^n)=(-1)^{k_n} H_{k_n}(q^n)$ and the multipaths $\pm q^n({\bf x})$ of the wave eigenfunctions $\Psi_{k_n}(q^n)$ of (\ref{eq:HermitePsi}).

Performing with (\ref{eq:HermitePsi}) a Hilbert space decomposition \cite{hilbert1904} of the initial classical density distribution
$\sqrt{\rho_j}({\bf x}_o, 0) =  \sum_{k \in \mathbb{N}^\star}^{\forall k_1+ ...+ k_4 = 2k-2} c_{k_1 .. k_N} \prod_{n=1}^N  \ \Psi_{k_n}(x^n_o)$ , the wave (\ref{eq:eigenstatexxo}) can be rewritten
\begin{eqnarray}
\psi({\bf q}, t) &=&  \sum_{k \in \mathbb{N}^\star}^{\forall k_1+ ...+ k_4 = 2k-2} c_{k_1 .. k_4} \ e^{\frac{i}{\hbar} E_k t} \prod_{n=1}^4 \Psi_{k_n}(q^n)   \nonumber
\end{eqnarray} 

While computed just from the $\mathbb{J} = {\bf q}_o \in \mathbb{R}^4 \ \times \ \{ \{ \uparrow, \downarrow \} \ \times k \ \in \mathbb{N}^\star  \}$ classical counter-rotating Kepler orbits in Figure \ref{fig:Coulomb}, this result matches the 3-dimensional Coulomb wave in spherical coordinates \cite{Schrodinger1926Atom, Duru, zwiebach2022}. 

Figures \ref{fig:Coulomb} to \ref{fig:CoulombProb2} illustrate the classical Kepler orbits and quantum probability density (\ref{eq:Prob}, \ref{eq:qwave}) in Cartesian coordinates $x^2, x^3$ (\ref{eq:quaternion}) and time $t$ for $\sqrt{\frac{4 M \omega}{\hbar}} = 1, q^3=0, q^4 = 0, r = (q^1)^2 + (q^2)^2$. Note that different eigenfunctions of the same degenerate eigenvalue can be superimposed to generate new eigenfunctions of the same degenerate eigenvalue. Also, the superposition principle applies both to the Coulomb wave (\ref{eq:qwave}), and to the Kepler orbits, whose position dynamics (\ref{eq:HJBGravity}) is linear in ${\bf q}, t'$ coordinates.
\begin{itemize}
    \item Figure \ref{fig:Coulomb} shows two Kepler orbit pairs with starting position $(q_o^1, q_o^2) = (-5, -5)$ or $(0, 5)$, which is in Cartesian coordinates $(x^2_o, x_o^3) = (-50, 0)$ or $(0, 25)$, and end position $(q^1, q^2) = \pm (3, 2)$, which is in Cartesian coordinates $(x^2, x^3) = (-12, -5)$. The two counter rotation solutions correspond to $\pm {\bf q}$.
\item Figure \ref{fig:CoulombProb0} shows the quantum eigendensity $\varrho = \psi \psi^{\dagger} = \frac{1}{\sqrt{\pi} 2^{0} 0!} (H_o(q^1) H_o (q^2))^2 e^{- \frac{r}{2}} = \frac{1}{\sqrt{\pi}}e^{- \frac{r}{2}}$  for $k=1, c_{k_1=k_2=0}=1$ of (\ref{eq:qwave}), called orbit 1S.  
\item Figure \ref{fig:CoulombProb} shows the quantum eigendensity $\varrho = \psi \psi^{\dagger} = \frac{1}{\sqrt{\pi} 2^1 1!} (H_1(q^1) H_1 (q^2))^2 e^{- \frac{r}{2}} = \frac{1}{2 \sqrt{\pi}}  (4 q^1 q^2)^2 e^{- \frac{r}{2}} = \frac{2}{\sqrt{\pi}}  (x^2)^2 e^{- \frac{r}{2}}$ for $k=2, c_{k_1=k_2=1}=1$ of (\ref{eq:qwave}), called orbit 2P. The 2P orbit can be superimposed to the 2S orbit, which is the 2P orbit rotated by 90° with the same eigenvalue. 
\item Figure \ref{fig:CoulombProb2} shows the quantum eigendensity $\varrho = \psi \psi^{\dagger} = \frac{1}{\sqrt{\pi( 2^1 1!)(2^{3} 3!)}} (H_1(q^1) H_3 (q^2)-H_3(q^1) H_1 (q^2))^2 e^{- \frac{r}{2}}$ for  $k= 3, c_{k_1=1, k_2=3}=1, c_{k_1=3, k_2=1}=-1$ of (\ref{eq:qwave}), called orbit 3D.  
\end{itemize}
There is no process noise added to the classical path in contrast to the Feynman path integral \cite{Feynman} in the Duru-Kleinert propagator \cite{Duru}. Hence the eigenwaves are now derived from the $\mathbb{J} = \{ {\bf q}_o \in \mathbb{R}^4 \} \ \times \ \{ \{ \uparrow, \downarrow \} \ \times k \ \in \mathbb{N}^\star  \}$  determined Kepler orbits \cite{Kepler}. 

One of the motivations for Bohr's atom model in the early days of quantum mechanics was the instability of previous orbital models, where the magnetic field generated by the electron's circular motion led to continual radiation and thus collapse of the atom. This problem is avoided here, since the effects of the two counter-rotating magnetic fields cancel each other.}{Coulomb}

\section{The relativistic case} \label{relativity}

Lemma \ref{lem:equivalence} is actually a general mapping from a quadratic first-order p.d.e. to a linear second-order p.d.e., i.e., it is not restricted to the Hamilton-Jacobi or Schr\"odinger equations. Specifically, the following three sections will apply Lemma \ref{lem:equivalence} and Theorem \ref{th:quantum} to
\begin{itemize}
    \item The Klein-Gordon equation, by replacing the classical action $\phi$ with a relativistic action.
    \item The Dirac and Pauli equations, by replacing the classical action $\phi$ with a unit quaternion action. 
    \item The relativistic Maxwell equation, by replacing the classical action $\phi$ with a relativistic action with rest mass $M_o = 0$. 
\end{itemize}

In general relativity, the metric tensor ${\bf M}'$ is defined by the Einstein field equation (EFE) \cite{Einstein}. The metric has a direct impact on the action and therefore on the wave in (\ref{eq:equivalence}). As such the EFE, which includes gravity, is implicitly covered in the three relativistic cases above. In special relativity, the metric is the  Minkowski metric 
\begin{equation}
   {\bf M}  = \ M_o \ {\bf M}' \ \ \ \ \ \ \ \ \ \ \ \ \ \ \ \ \ \ {\bf M}' = \ {\rm diag}(RI^2 c^2, -1, -1, -1) \label{eq:Minkowski}
\end{equation}
with constant rest mass $M_o \ $, speed of light constant $c \ $, and refractive index $RI \le 1$ (where $RI=1$ in  free space). We use ${\bf M}$ for massive particles and ${\bf M}'$ for massless particles.

\subsection{Klein-Gordon equation} \label{KleinGordon}

The Hamilton-Jacobi p.d.e. (\ref{eq:HJ}) also applies to general relativity~\cite{Einstein, Liboff} when we replace 
\begin{eqnarray}
{\bf x} & \rightarrow  & \bar{\bf x} = \left( \begin{array}{c}
          t \\
        {\bf x}
    \end{array} \right),    
\ {\bf A} \rightarrow  \bar{\bf A} = \left( \begin{array}{c}
        V_E (\bar{\bf x}) \\
        {\bf A}(\bar{\bf x})
    \end{array} \right),     
\  {\bf p} = \frac{\partial \phi}{\partial {\bf x}} \rightarrow  \bar{\bf p} = \frac{\partial \phi}{\partial \bar{\bf x}} = \left( \begin{array}{c}
        E (\bar{\bf x}) \\
        {\bf p}(\bar{\bf x})
    \end{array} \right) \ \ \ \ \ \ \ \label{eq:relativity} \\ 
     t &\rightarrow&  \tau = \int \sqrt{ \frac{d \bar{\bf x}}{d t}^T  {\bf M}'(\bar{\bf x} ) \frac{d \bar{\bf x}\ } {d t}}\ \frac{d t}{c}  \ \ge \ 0  \nonumber
\end{eqnarray}
with classical momentum ${\bf p}$, relativistic total energy $E$, electrostatic potential $V_E \ $, charge $Q$, and proper time $\tau$ path integral, measured by a clock attached to the particle. In proper time, i.e., in a coordinate frame attached to the particle, the energy of a particle is always the constant rest energy $E_o =M_o c^2\ $. Hence, the action in Theorem \ref{th:Hamilton} and the wave in Theorem \ref{th:quantum} transform as 
\begin{eqnarray}
    \phi({\bf x}, t) \ \rightarrow \  \Phi(\bar{\bf x}) \ \ \ \ \ \ \ \ \ \  
    \psi({\bf x}, t) \ \rightarrow  \ \psi(\bar{\bf x}) 
    \ \ \ \ \ \ \ \ \ \ \bar{\bf x} \in \mathbb{G}^4 \nonumber
\end{eqnarray} 
 We thus have the relativistic covariant Hamilton-Jacobi p.d.e. for a single particle, otherwise known as the relativistic energy momentum relation \cite{Einstein:1905,  Einstein}, for $j = 1, ..., J$
\begin{eqnarray}
 - \frac{\partial \phi_j}{\partial \tau} \ = \ 0 \ = \ H  &=&  
\frac{1}{2 }\frac{d \bar{\bf x}}{d \tau}^T  {\bf M}  \frac{d \bar{\bf x}}{d \tau} - \frac{E_o}{2} \label{eq:relHJ} \\
   \frac{d \nabla \phi_j}{d \tau}   +   \frac{\partial H}  {\partial \bar{\bf x}}\  &=&  \sum_{g \in \mathbb{G}} \frac{\partial f_g}{\partial \bar{\bf x}} \lambda_g,   \ \ \ 
   {\bf M} \ \frac{d \bar{\bf x}}{d \tau} \  = \ \nabla \phi_j - Q \bar{\bf A} \nonumber
\end{eqnarray} 
Note that in standard derivations of the relativistic quantum equations \cite{Klein, Gordon,Dirac28,zwiebach2022}, the Hamiltonian is set to $E$ and not to $H$ in (\ref{eq:relHJ}). This prevents a direct generalization of the Schr\"odinger equation (\ref{eq:Schroed}) to the relativistic case, as we do here and in the next subsections. 
The relativistic density is given by the relativistic continuity equation 
\begin{eqnarray}
0 \ = \ \frac{\partial}{\partial \tau} \ \rho_j + \nabla_{\bf M} \cdot (\rho_j \ \dot{\bar{\bx}} ) &=&   \frac{d \rho_j }{d\tau} \ + \rho_j \ \nabla_{\bf M} \cdot \dot{\bar{\bx}}  \label{eq:densityrel}
\end{eqnarray} 
whose solution always remains positive. Using Lemma \ref{lem:equivalence} yields the familiar Klein-Gordon equation \cite{Gordon, Klein, Liboff}
\begin{eqnarray}
    0 &=&  \left[ \left( \frac{\hbar}{i} \nabla_{{\bf M}} - Q \bar{\bf A} \right) \cdot {\bf M}^{-1} \left( \frac{\hbar}{i} \nabla - Q \bar{\bf A} \right) - E_o   \right]  \Psi (\bar{\bf x}) \label{eq:KleinGordon}
\end{eqnarray}
Hence the Klein-Gordon equation is 
a special case of Theorem \ref{th:Hamilton} and Theorem \ref{th:quantum} for unconstrained $\bar{\bf x} \in \mathbb{R}^4$. Note that Theorem \ref{th:quantum} can be seen as an extension of the non-relativistic zig-zag Feynman path integral (\ref{eq:Feynman}) of \cite{Feynman48} to the relativistic case, respecting the speed of light limit. 

Importantly, the proper density (\ref{eq:densityrel}) is defined  with respect to a proper $4$-dimensional volume element, which does not Lorentz contract \cite{Lorentz1904, Einstein:1905}, by contrast to a $3$-dimensional volume. Hence the quantum density matrix  $\varrho  =  \psi \psi^{\dagger}$ of (\ref{eq:Prob}) is the coordinate invariant positive probability to find a particle in a $4$-dimensional volume element. Replacing $t$ by $\tau$ in (\ref{eq:prob1}), $\varrho$ normalized at $\ \tau = \tau_o \ $ remains normalized $ \forall \tau \ge \tau_o \ $.

\subsection{Dirac and Pauli equations} \label{Dirac} 

Quaternion rotation of objects~\cite{Goldstein} is well defined in geometry, and is used in an identical form in classical, relativistic, and quantum physics. This section defines accordingly a quaternion-based rotation action, and shows how the Dirac and Pauli equations can be derived from this classical quaternion action.
We use standard Pauli spin \cite{Pauli} and Dirac matrices \cite{Dirac28, feynman_quantum_1998}, 
\begin{eqnarray}
     {\bf \Sigma} &=& (\sigma^1, \sigma^2, \sigma^3) \ \ \ \ \ \
     \sigma^1 = \left( \begin{array}{cc}
        0 & 1 \\
        1 & 0
    \end{array} \right) \ \ \ \ \ \   \sigma^2 = \left( \begin{array}{cc}
        0 & -i \\
        i & 0
    \end{array} \right) \ \ \ \ \ \  \sigma^3 = \left( \begin{array}{cc}
        1 & 0 \\
        0 & -1
    \end{array} \right) \ \ \ \ \label{eq:Sigma}  \\
    {\bf \Gamma} &=& (\gamma^0,\gamma^1, \gamma^2, \gamma^3)\ \ \ \
        \gamma^o = \left( \begin{array}{cc}
        {\bf I} & {\bf 0} \\
        {\bf 0} & - {\bf I}
    \end{array} \right)  \ \ \ \  
     \gamma^n =  \left( \begin{array}{cc}
        {\bf 0} & \sigma^n \\
        \sigma^n & {\bf 0}
    \end{array} \right) \  {\rm for} \  \ n =1, 2, 3  \ \   \label{eq:Gamma} 
\end{eqnarray}

\begin{figure}
\centering
    \includegraphics[scale=0.45]{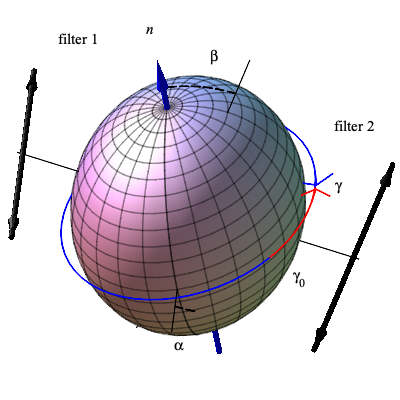}
    \caption{Two classical counter rotations} 
    \label{fig:spin}
\end{figure}

In the non-relativistic case, we replace the scalar action in Theorem \ref{th:Hamilton} with a $2 \times 2$ pure imaginary quaternion action, which implies a unit quaternion wave in Lemma \ref{lem:equivalence}, 
\begin{eqnarray}
\phi  \rightarrow  {\bf \phi} &=& \hbar \ {\bf \Sigma} \cdot {\bf n} \ \frac{s \gamma}{2} \in \mathbb{H} \label{eq:purequat} \\
\psi \rightarrow  {\bf \psi} &=& {\bf I} \cos{\frac{s \gamma}{2}} + i  {\bf \Sigma} \cdot {\bf n} \sin{\frac{s \gamma}{2}} = \ e^{i  {\bf \Sigma} \cdot {\bf n}  \frac{s \gamma}{2}} \ \ \in \ \mathbb{H}_1  \label{eq:H1} 
\end{eqnarray}
with the roll angle rotation $ \ - \pi \le s \gamma \le \pi \ $  in Figure \ref{fig:spin} around the unit direction ${\bf n} = (n^1, n^2, n^3)^T = (\sin{\beta} \cos{\alpha}, \sin{\beta} \sin{\alpha}, \cos{\beta})^T$, with Euler yaw $- \pi \le \alpha \le \pi$ and pitch $0 \le \beta \le \pi$. The particles are rotationally symmetric of order $\frac{2 \pi}{s}$ \cite{Hawking1975}, with $s \in \mathbb{N}^\star$.
Fermions and bosons correspond to $s=1$ and $s=2$. The $2 \times 2$ unit quaternion (\ref{eq:H1}) can be modally decomposed as
\begin{eqnarray}
{\bf \Sigma} \cdot {\bf n}  &=  \left( \begin{array}{cc}
        \cos \beta & e^{-i \alpha}\sin \beta  \\ 
        e^{i \alpha} \sin \beta & - \cos \beta 
    \end{array} \right) &= \ {\bf \chi}^{\uparrow}  ({\bf \chi}^{\uparrow})^{\dagger}   - {\bf \chi}^{\downarrow} ({\bf \chi}^{\downarrow})^{\dagger} \label{eq:sigman}  \\
    {\bf \chi}^{\uparrow} &= \left( \begin{array}{c}
        \cos \frac{\beta}{2}  \\ 
       e^{i \alpha} \ \sin \frac{\beta}{2}
    \end{array} \right) \ \ \ \ \ \ \ \ \ \ \ \ 
    {\bf \chi}^{\downarrow} &= \left( \begin{array}{c}
        -e^{- i \alpha}\ \sin \frac{\beta}{2}  \\ 
        \cos \frac{\beta}{2} 
    \end{array} \right) \label{eq:eigenspinor}
\end{eqnarray}
The orthonormal eigenvectors ${\bf \chi}^{\uparrow}({\bf n})$ and ${\bf \chi}^{\downarrow}({\bf n}) \ $ of $ \ {\bf \Sigma} \cdot {\bf n}$, called eigenspinors~\cite{Goldstein}, correspond to an aligned $\uparrow$ or anti-aligned $\downarrow$ classical rotation around ${\bf n}$ in Figure \ref{fig:spin} of the Bloch sphere \cite{Bloch1946NuclearInduction}. 
In contrast to Euler angles, unit quaternions yield a linear motion description and hence allow a simple application of Theorems \ref{th:Hamilton} and \ref{th:quantum}. Note that translational and rotational dynamics are uncoupled (in the absence of external coupling forces).

In the relativistic case, we replace instead the scalar action in Theorem \ref{th:Hamilton} with a pure imaginary $4 \times 4$  quaternion action,  which also implies a unit quaternion wave in Lemma \ref{lem:equivalence},
\begin{eqnarray}
\phi \ \rightarrow \  {\bf \phi}  &=& \hbar \ {\bf \Gamma} \cdot \bar{\bf p} \  \frac{s \gamma}{2} \ \in \ \mathbb{H} \label{eq:purequatrel} \\
\psi \ \rightarrow \  {\bf \psi}  &=&    {\bf I} \ \cos{\frac{s \gamma}{2}} + \frac{i}{\hbar}   {\bf \Gamma} \cdot \bar{\bf p} \ \sin{\frac{s \gamma}{2}} = \ e^{\frac{i}{\hbar}  {\bf \Gamma} \cdot \bar{\bf p}   \frac{s \gamma}{2}} \ \ \in \ \mathbb{H}_1  \label{eq:H1Rel} 
\end{eqnarray}
Both wave quaternions are differentiable for constant ${\bf n}$ and $\bar{\bf p}$. The $4 \times 4$ quaternions (\ref{eq:purequatrel}) can be modally decomposed as  \cite{Dirac28, Liboff}
\begin{eqnarray}
c \ {\bf \Gamma} (\alpha_p, \beta_p) \cdot \bar{\bf p}  
&=&  E_+ {\bf \xi}_{\uparrow}^+ {\bf \xi}_{\uparrow}^{+\dagger} + E_+ {\bf \xi}^+_{\downarrow} {\bf \xi}_{\downarrow}^{+\dagger} + E_- {\bf \xi}_{\uparrow}^- {\bf \xi}_{\uparrow}^{-\dagger} + E_- {\bf \xi}^-_{\downarrow} {\bf \xi}_{\downarrow}^{-\dagger}  \label{eq:chidownrel}
\end{eqnarray} 
where the ambiguity of having $4$ relativistic eigenspinors (\ref{eq:xi}) for only $2$ eigen-energies is resolved by using the classical eigenspinors ${\bf \chi}^{\downarrow, \uparrow}(\alpha_p, \beta_p)$ of  (\ref{eq:eigenspinor}) (see Figure \ref{fig:spin}), 
\begin{eqnarray}
    {\bf \xi}_{\uparrow, \downarrow}^+ &=& \frac{1}{N_+} \left( \begin{array}{cc}
        {\bf I}  \\ 
        {\bf \Sigma}  \cdot {\bf p} 
    \end{array} \right) 
    \ {\bf \chi}^{\uparrow, \downarrow} \ \ \ \ \ \ \ \ \ \ \ \  
    {\bf \xi}_{\uparrow, \downarrow}^- = \frac{1}{N_+} \left( \begin{array}{cc}
        {\bf \Sigma}  \cdot {\bf p} \\ 
         {\bf I}  
    \end{array} \right) 
     \ {\bf \chi}^{\uparrow, \downarrow} \label{eq:xi} \\
    E_{\pm} &=& \pm \sqrt{E_o^2  + {\bf p}^T {\bf p} \ c^2} \ \ \ \ \ \ \ \ \ \ \ \Delta_{\pm} = E_{\pm} \pm E_o \ \ \ \ \ \ \ \ \ \ \ N_{\pm} = \sqrt{1+ \frac{c^2 {\bf p}^T {\bf p}}{\Delta_{\pm}^2}} \nonumber
\end{eqnarray} 
The orthonormal eigenvectors ${\bf \xi}_{\uparrow, \downarrow}^+, {\bf \xi}_{\uparrow, \downarrow}^-$, called eigenspinors~\cite{Goldstein}, correspond to an aligned $\uparrow$ or anti-aligned $\downarrow$ relativistic rotation around ${\bf n}$ in Figure \ref{fig:spin} for a positive or negative eigen-energy $E_{\pm} \ $. Note for massless particles the $4 \times 1$ Dirac spinor (\ref{eq:xi}) simplifies into two decoupled $2 \times 1$ complex chiral spinors.

The actions and waves (\ref{eq:purequat}) - (\ref{eq:H1Rel})  are quaternions due to the anti-commutation (decoupling) relation 
\begin{eqnarray}
       \{ \sigma^{j}, \sigma^{k}\} &=& \sigma^{j}  \sigma^{k} + \sigma^{k} \sigma^{j} = 2 \ \delta^{jk} 
    \label{eq:anticommute} \\ 
        \{\gamma^{j}, \gamma^{k}\} &=& \gamma^{j}\gamma^{k} + \gamma^{k} \gamma^{j} = 2 \ \delta^{jk}  \label{eq:anticommuteRel} 
\end{eqnarray}
 Using Lemma \ref{lem:equivalence}, the classical action (\ref{eq:purequat}) now implies the Pauli equation
\begin{equation}
\left[ \frac{\hbar}{i} \frac{\partial}{\partial t} + \frac{1}{2} {\bf \Sigma} \cdot \left( \frac{\hbar}{i} \nabla - Q {\bf A}  \right)  \cdot \ {\bf M}^{-1} {\bf \Sigma} \cdot \left( \frac{\hbar}{i} \nabla - Q {\bf A}  \right) +V \right] {\bf \psi} = 0 \label{eq:Pauli}
\end{equation}
The scalar wave function $\psi$ is replaced by a $2 \times 1$ complex spinor (\ref{eq:eigenspinor})


Still using Lemma \ref{lem:equivalence}, the relativistic action (\ref{eq:purequatrel}) now implies the Dirac equation 
\begin{eqnarray}
{\bf 0} &=& \left[ {\bf \Gamma} \cdot \left( \frac{\hbar}{i} \nabla - Q \bar{\bf A}  \right) + M_o c  \right] {\bf \psi} \label{eq:DiracEq} 
\end{eqnarray}
multiplied from the left with the invertible matrix $\left[ {\bf \Gamma} \cdot \left( \frac{\hbar}{i} \nabla - Q \bar{\bf A}  \right) - M_o c  \right] \cdot \ {\bf M}^{-1}$. We use the $4 \times 1$ relativistic spinor (\ref{eq:xi}).
Hence the Pauli or Dirac equation are for $\phi \in \ \mathbb{R} \rightarrow  {\bf \phi} \in \ \mathbb{H}$ special cases of Theorem \ref{th:Hamilton} and Theorem \ref{th:quantum}. 

\subsection{Maxwell equation} \label{Maxwell}

The action and motion of a photon are given relativistically by (\ref{eq:relHJ}), with zero rest mass $M_o = 0$, zero rest energy $E_o = 0$ and without external forces~\cite{Einstein}. We use the relativistic metric ${\bf M}'$ of (\ref{eq:Minkowski}) instead of the relativistic inertia tensor ${\bf M}$. Hence Theorem \ref{th:Hamilton} and (\ref{eq:relHJ}) yields the relativistic eikonal equation of geometric ray optics \cite{hamilton1828rays}, 
\begin{eqnarray}
0 \ = \ H \ = \  \frac{1}{2} \ \nabla {\Phi}_j^T \ {\bf M}^{'-1} \nabla {\Phi}_j \ \ \ \label{eq:MaxAction} 
\end{eqnarray}
i.e., a photon moves along a geodesic of the relativistic metric ${\bf M}^{'}$ of Einstein's field equation (EFE) \cite{Einstein}. 
We now replace in Lemma \ref{lem:equivalence} the wave of a massive particle with the real relativistic $4 \times 1$ vector potential $\bar{\bf A}'$  of the photon
\begin{eqnarray}
\Psi &\rightarrow& \bar{\bf A}' = \sqrt{\rho_j} \ e^{\frac{i \Phi_j}{\hbar}} \nonumber 
\end{eqnarray}
where $\sqrt{\rho_j}$ is a $4 \times 1$ square root density vector. 
Now Lemma \ref{lem:equivalence} yields the familiar homogeneous Maxwell equation \cite{Maxwell1865} 
\begin{eqnarray}
  \Delta_{{\bf M}'} \ \bar{\bf A}' \ = \  {\bf 0} \label{eq:HomMaxwell}  
\end{eqnarray}
which are four decoupled Laplacians. So far the norm $|\bar{\bf A}'|$ of the vector potential has remained constant, and hence photons cannot be created nor annihilated in (\ref{eq:MaxAction}) or (\ref{eq:HomMaxwell}) by changing the photon density $\sqrt{\rho_j} = {\bf 0}$ to $\sqrt{\rho_j} \ne {\bf 0}$, or vice-versa. This is not surprising, since (\ref{eq:MaxAction}) is just the geodesic path of a single photon. Photon creation and annihilation can be obtained by replacing  (\ref{eq:HomMaxwell}) with the non-homogeneous Maxwell equation 
\begin{eqnarray}
  \Delta_{{\bf M}'} \ \bar{\bf A} \ = \   \frac {1}{\mu_{0}} \ \bar{\bf J}  \label{eq:Maxwell}  
\end{eqnarray}
where $\mu_o$ is the magnetic permeability and $\bar{\bf J}(\bar{\bf x})$ is the given relativistic current density. Since Maxwell's homogeneous equation (\ref{eq:HomMaxwell}) is linear, (\ref{eq:Maxwell}) can be solved by convolution,
\begin{equation}
    \bar{\bf A} (\bar{\bf x}) = \frac {1}{\mu_{0}} \int_{\bar{\bf z} \in \mathbb{G}^N} \bar{\bf A}^{'}(\bar{\bf x}-\bar{\bf z}) \ \bar{\bf J}(\bar{\bf z}) \ d \bar{\bf z} \nonumber
\end{equation}
where all products are performed element-wise. Thus $\frac {1}{\mu_{0}} \ \bar{\bf J}$ changes the norm $|\bar{\bf A}|$, which implies the creation and annihilation of photons. Hence, with $\psi \in \ \mathbb{R} \ $ replaced by  $ \ \bar{\bf A} \in \ \mathbb{R}^4$, Maxwell's equation is also a special case of Theorem \ref{th:Hamilton} and Theorem \ref{th:quantum}. 

\subsection{Examples with multiple particles}

We now turn to examples involving multiple spinning particles, starting with the Einsten-Podolsky-Rosen~ experiment~\cite{Einstein-Podolsky,Bell1964,clauser1969, Aspect,zeilinger2000physics}. The fact that the quantum wave in the experiment can be computed using a finite sum of classically-based terms puts
its discussion in a different light. We first detail this computation, and then relate it to Bell's Theorem.

\Example{ EPR experiment and entanglement \ \ } {Consider $P$ spinning particles. Since there is no potential energy, the particles $p \in \mathbb{P} = \{ 1, ..., P \}$ are classically decoupled. This implies that the total Hamiltonian and action sum up as $\ {\bf H} = \sum_{p \in \mathbb{P}} {\bf H}_p\ $ and $\ \phi = \sum_{p \in \mathbb{P}} \phi_p \ $. We describe the rotation of each individual particle $p$ with a pure imaginary quaternion action (\ref{eq:purequat}) in the decoupled Hamilton-Jacobi p.d.e. (\ref{eq:HJ}),
\begin{eqnarray}
    -  \frac{\partial {\bf \phi}_p}{\partial t} = {\bf H}_p  \ = \frac{1}{2 M} \left( \nabla {\bf \phi}_p^T \nabla {\bf \phi}_p - \frac{s^2 \hbar^2}{4} {\bf I} \right) \ \ \ \ \ \ \ \ \ \ \ \ \ \ \ \ \ {\bf \phi}_p \ \  \in \mathbb{H} \nonumber 
\end{eqnarray}
with the classical eigenspinors ${\bf \chi}_{p}^{\uparrow, \downarrow}({\bf n}_{p})$ (\ref{eq:eigenspinor}) of unit rotation direction ${\bf n}_p(\alpha_{p}, \beta_{p})$ and roll or spin angle $0 \le \gamma_p \le \frac{2 \pi}{s}$ in Figure \ref{fig:spin}. The initial rotation direction is given by the initial classical eigenspinors  ${\bf \chi}_{po}^{\uparrow, \downarrow}({\bf n}_{po})$ of the initial unit rotation direction $ {\bf n}_{po}(\alpha_{po}, \beta_{po})$ of (\ref{eq:eigenspinor}). The particles have the same moment of inertia $M$ and they  are rotationally symmetric of the same order $\frac{2 \pi}{s}$, with $s \in \mathbb{N}^\star$.

The $ \ \mathbb{J} =  \{  \gamma_{po} \in [0, \frac{2 \pi}{s}] + \gamma_p  \} \ \times \ \{\uparrow, \downarrow \}$ - valued least quaternion action (\ref{eq:purequat}) of each particle $p$ in Theorem \ref{th:Hamilton} is
\begin{equation}
\frac{1}{\hbar }{\bf \phi}_{pj}({\bf n}_{p}, \gamma_p) \ =    \ {\bf \Sigma} \cdot {\bf n}_p \left\{ 
\begin{array}{ll}
 \frac{s}{2} (\gamma_p - \gamma_{po})  & \ \text{\ \ \ \ \ for \ }  \uparrow \\
 \frac{s}{2} (\frac{2 \pi}{s}  - (\gamma_p - \gamma_{po}))  & \ \text{\ \ \ \ \  for \ }  \downarrow 
\end{array}
\right. \in \mathbb{H}\label{eq:quataction}
\end{equation} 
where we exploited the anti-commutation relation (\ref{eq:anticommute}) of the Pauli matrices (\ref{eq:Sigma}). While the Hamiltonian  ${\bf H}_p$ is zero, the action is non-zero and thus the particle can rotate to the left and right in Figure \ref{fig:spin} with constant spin momentum $\nabla {\bf \phi}_{pj} = \pm \frac{s}{2} \hbar {\bf \Sigma} \cdot {\bf n}_p \ $. 

Since $\Delta_M {\bf \phi}_{pj} = {\bf 0}$ in (\ref{eq:density}), an initial classical Dirac spinor distribution remains a Dirac density distribution  for all $t \ge 0$, i.e., we can define the spinor density as $4 i \sqrt{\rho_{p j}}^{\epsilon}  = {\bf \chi}_{p}^{\epsilon} = {\bf \chi}_{p o}^{\epsilon}$ with $ \epsilon \in \mathbb{E}_p = \{ \uparrow, \downarrow \}$.

Thus, from  Theorem \ref{th:quantum} and (\ref{eq:sigman}), the wave spinor (\ref{eq:eigenspinor}) of particle $p$,
\begin{equation}
{\bf \psi}_{p}^{\epsilon} ({\bf n}_p, {\bf \chi}_{po}^{\epsilon}) = \sum_{j \in \mathbb{J}}  \sqrt{\rho_{pj}}^{\epsilon} \ \ e^{\frac{i }{\hbar} {\bf \phi}_{p j}}  
= \frac{1}{2} \int_{\gamma_{po} = \gamma_p}^{\gamma_p+ \frac{2 \pi}{s} } \sin{\frac{s}{2} (\gamma_p - \gamma_{po})}  d \gamma_{po}  {\bf \Sigma} \cdot {\bf n}_p  {\bf \chi}_{po}^{\epsilon}  =  {\bf \Sigma} \cdot {\bf n}_p  {\bf \chi}_{po}^{\epsilon} \ \ \ \label{eq:sigmasigma0} 
\end{equation}
solves the Pauli equation (\ref{eq:Pauli}) with every summand $\sqrt{\rho_{p j}}^{\epsilon}  \ e^{\frac{i }{\hbar} {\bf \phi}_{p j}}$. 
Equation (\ref{eq:sigmasigma0}) computes the wave spinor ${\bf \psi}_{p}^{\epsilon}$ 
\begin{itemize}
    \item  as the geometric projection of an initial classical spinor ${\bf \chi}_{po}^{\epsilon}$ on a filter of direction ${\bf n}_p (\alpha_p, \beta_p)$. The filter is a branch point of the action (\ref{eq:quataction}), since it changes the spin direction of the particle from ${\bf n}_{po}$ to ${\bf n}_{p}$ .
    \item  as a solution of the eigenspinor equation (\ref{eq:sigman}) of a pure left or right rotation for ${\bf \psi}_{p}^{\epsilon} = {\bf \chi}_{p}^{\epsilon} = {\bf \chi}_{po}^{\epsilon}$ in free space. A superposition of wave spinors ${\bf \psi}_{p}^{\epsilon}$ derives from a classical density ensemble (\ref{eq:density}) of initial ${\bf \chi}_{po}^{\epsilon} \ $.
\end{itemize}

Finally, aside from initial conditions, the total action $\phi = \sum_{p \in \mathbb{P}} \phi_{p}({\bf x}_p, t)$ consists of $P$ decoupled pure imaginary quaternion action fields. Hence  the corresponding total wave (\ref{eq:Wave}) consists of $P$ decoupled spinors, and thus may be represented as the tensor product of the individual spinors with $\epsilon \in \mathbb{E} = \{ \mathbb{E}_1, .. , \mathbb{E}_P  \}$
\begin{eqnarray}
   {\bf \psi}^{\epsilon} ({\bf n}_p, {\bf \chi}_{po}^{\epsilon}) &=& \sum_{p \in \mathbb{P}} {\bf \psi}^{\epsilon}_1 \otimes ... \otimes {\bf \psi}^{\epsilon}_{P} \ = \sum_{p \in \mathbb{P}} \ {\bf \Sigma} \cdot {\bf n}_1 \ {\bf \chi}_{1o}^{\epsilon} \otimes ... \otimes {\bf \Sigma} \cdot {\bf n}_P \ {\bf \chi}_{Po}^{\epsilon} \label{eq:entanglement}
\end{eqnarray}

Consider now the Einstein-Podolsky-Rosen (EPR) experiment~\cite{Einstein-Podolsky,clauser1969, Aspect,zeilinger2000physics} with two particles $p=1, 2$ (each as in Figure \ref{fig:spin}) which initially have opposite spins in the two ensembles $\mathbb{E} = \{ \uparrow \downarrow, \downarrow \uparrow \}$ of probability $p^{\epsilon} = \frac{1}{2}$. 
Both ensembles have a classical total spin of $0$. Thus the spin of the particles is ${\bf \psi}_o = \frac{1}{\sqrt{2}}\left( {\bf \psi}^{\uparrow}_{o} \otimes {\bf \psi}^{\downarrow}_{o} - {\bf \psi}^{\downarrow}_{o} \otimes {\bf \psi}^{\uparrow}_{o} \right)$ with an unknown initial spin direction ${\bf n}_o(\alpha_o, \beta_o)$. Later on, particle $1$ is measured behind a filter $1$ with angles ${\bf n}_1 (\alpha_1, \beta_1)$ and at a far distance particle $2$ is measured behind a filter $2$ with angles ${\bf n}_2 (\alpha_2, \beta_2)$ in Figure \ref{fig:spin}. 
With (\ref{eq:entanglement}) this yields
\begin{eqnarray}
   {\bf \psi} &=& \frac{1}{\sqrt{2}}\left( {\bf \psi}^{\uparrow}_{1} \otimes {\bf \psi}^{\downarrow}_{2} - {\bf \psi}^{\downarrow}_{1} \otimes {\bf \psi}^{\uparrow}_{2} \right) \ \ \ \ \ \rm{where} \ \ \ \ {\bf \psi}^{\uparrow}_{p} = {\bf \Sigma} \cdot {\bf n}_p \ {\bf \chi}^{\uparrow}_{o} \ \ , \ \ {\bf \psi}^{\downarrow}_{p} =  {\bf \Sigma} \cdot {\bf n}_p \ {\bf \chi}^{\downarrow}_{o} \ \ \ \ \ \ \ \ \ \ \ \label{eq:spinEPR}
\end{eqnarray}
The two spin measurements are the geometric projections of an unknown initial spin direction $\ {\bf \chi}^{\uparrow}_{o} (\alpha_o, \beta_o)$ , ${\bf \chi}^{\downarrow}_{o} (\alpha_o, \beta_o) \ $ on filters of directions ${\bf \Sigma} \cdot {\bf n}_1$ and ${\bf \Sigma} \cdot {\bf n}_2$. 
The correlation of the two measurements is given by the scalar product of the spinor measurements 
\begin{equation}
\langle {\bf \psi}_1^{\uparrow}\ ,  \ {\bf \psi}_2^{\downarrow} \rangle \ \ = \ \frac{1}{2} \left( {\bf \psi}_1^{\uparrow \dagger} {\bf \psi}^{\downarrow}_2 + {\bf \psi}^{\downarrow \dagger}_2 {\bf \psi}^{\uparrow}_1 \right) \ 
=  \ \langle {\bf \Sigma} \cdot {\bf n}_1 \ {\bf \chi}^{\uparrow}_{o}\ , \ {\bf \Sigma} \cdot {\bf n}_2 \ {\bf \chi}^{\downarrow}_{o} \rangle \ = \ - \ {\bf n}_1^T {\bf n}_2 \label{eq:EPRcor}
\end{equation} 
This expression is consistent with experimental results~\cite{Aspect}. The second ensemble $\  {\bf \psi}^{\downarrow}_{1} \otimes {\bf \psi}^{\uparrow}_{2} \ $ has the same correlation. 

For a free particle, the Hamilton-Jacobi formulation of Theorem \ref{th:Hamilton}, as well as Feynman's momentum propagator formulation in \cite{Feynman}, determine a constant classical linear or angular momentum ${\bf \chi}^{\uparrow}_{o}, {\bf \chi}^{\downarrow}_{o} \ $ over time.  Equation (\ref{eq:EPRcor}) thus allows a classical interpretation that the spins in the distant detectors are correlated through the common initial classical spinors ${\bf \chi}^{\uparrow}_{o}, {\bf \chi}^{\downarrow}_{o} \ $.
}{spin}

Let us put this result in historical perspective. Bell~\cite{Bell1964} derived the correlation of two classical spinning particles as the integral over a probability density function $\rho (\lambda ) \in \mathbb{R}$ of a hypothetical hidden parameter $\lambda \in \ \mathbb{R}$,
\begin{equation}
   P({\bf n}_1, {\bf n}_2) \ = \ \int  \rho (\lambda )A_1({\bf n}_1,\lambda ) A_2({\bf n}_2,\lambda )\ d\lambda \ \ \ \ \ \ \ \rm{with}\ \ \int  \rho (\lambda )\ d\lambda \ = \ 1 \label{eq:Bell}
\end{equation}
The spins are measured behind unit filters $p=1, 2 $ of direction ${\bf n}_p(\alpha_p, \beta_p)$ with the quantized detectors $ \ A_p({\bf n}_p,\lambda ) = \pm 1\ $, or equivalently
\begin{equation}
A_p({\bf n}_p,\lambda )  = {\bf \chi}^{\uparrow}_p, {\bf \chi}^{\downarrow}_p  \ \ \ \ \ \rm{where} \ \ \ {\bf \chi}^{\uparrow}_p = \left( \begin{array}{c}
        1 \\ 
        0 
    \end{array} \right), \ \  {\bf \chi}^{\downarrow}_p =  \left( \begin{array}{c}
        0 \\ 
        1 
    \end{array} \right) \ \ \label{eq:AB1}
\end{equation}
leading to Bell's inequality \cite{Bell1964} 
\begin{equation}
|P({\bf n}_1,{\bf n}_2)-P({\bf n}_1,{\bf n}_3)| \leq 1+P({\bf n}_2,{\bf n}_3) \label{eq:Bellin}
\end{equation}
with a third spin direction $p=3$. The correctness of (\ref{eq:EPRcor}) and incorrectness of (\ref{eq:Bellin}) were experimentally demonstrated in \cite{Aspect,clauser1969}. Hence it was argued that an exact classical or relativistic derivation of (\ref{eq:EPRcor}) is impossible since it violates (\ref{eq:Bellin}).


Spinors or eigenspinors (\ref{eq:eigenspinor}) on a unit Bloch sphere are exact quaternion descriptions of both classical rotations ${\bf \chi}_{p}^{\epsilon}$  and quantum rotations ${\bf \psi}_{p}^{\epsilon}$. Although the {\it classical  Bell detector} (\ref{eq:AB1}) depends on the local filter direction ${\bf n}_p \ $, the quantization is binary and corresponds to relative Euler angles $\beta = 2 k \pi, k \in \mathbb{Z}$ on the unit Bloch sphere. Generalizing the quantization of this classical binary Bell detector (\ref{eq:AB1}) to the full unit Bloch sphere allows to use the  {\it classical spinor detector} pair (\ref{eq:spinEPR}), 
\begin{eqnarray}
A_1 =  {\bf \Sigma} \cdot {\bf n}_1 \ {\bf \chi}^{\downarrow}_{o}\ , \ \ A_2 =  {\bf \Sigma} \cdot {\bf n}_2 \ {\bf \chi}^{\uparrow}_{o}  \label{eq:BlochDetector}
\end{eqnarray}
Using in Bell's hidden parameter equation (\ref{eq:Bell}) the general classical spinor detector pair
 (\ref{eq:BlochDetector}) in the directions of the filters,  instead of the classical binary Bell detector (\ref{eq:AB1}),  directly leads to (\ref{eq:EPRcor}) rather than (\ref{eq:Bellin}). Hence the classical correlation of two spinning particles is exactly described by (\ref{eq:EPRcor}), whereas Bell's inequality (\ref{eq:Bellin}) is classically only applicable at a relative angle $ \ \beta = 2 k \pi, k \in \mathbb{Z} \ $ on the Bloch sphere. For each particle, what is actually measured of course is just the up/down spin direction behind the filter, not the location on the Bloch sphere. However, the measurement result does depend on the filter, and the classical spinor detector is a description of the {\it combination} of the filter direction and the spin detector for the particle.  


The next example analyzes relativistic spin similarly.
\Example{Positron and electron creation in quantum electrodynamics (QED)}{ \ \ Consider $P$ relativistically spinning particles. The total Hamiltonian and action sums up as ${\bf H} = \sum_{p \in \mathbb{P}} {\bf H}_p, \phi = \sum_{p \in \mathbb{P}} \phi_p$. We describe the rotation of each individual particle $p$ with a pure imaginary quaternion action (\ref{eq:purequatrel}) in the decoupled Hamilton-Jacobi p.d.e. (\ref{eq:HJ}),
\begin{eqnarray}
    -  \frac{\partial {\bf \phi}_p}{\partial \tau} \ = \ {\bf H}_p  \ = \ \frac{1}{2} \left( \nabla {\bf \phi}_p {\bf M}^{-1} \nabla {\bf \phi}_p - \frac{s^2 \hbar^2 E_o^2}{4} {\bf I} \right) \ = \ {\bf 0} \ \ \ \ \ \ \ \ \ \ \ \ \ \ \ \ \ \ \ {\bf \phi}_p \ \  \in \mathbb{H}\nonumber 
\end{eqnarray}
with Minkowski metric ${\bf M}$ (\ref{eq:Minkowski}), rest energy $E_o\ $, constant relativistic linear momentum $\bar{\bf p}_{p} \ $ and the relativistic eigenspinors ${\bf \xi}_{p}^{\uparrow, \downarrow}({\bf n}_{p}, \bar{\bf p}_{p})$ (\ref{eq:xi}) of unit rotation direction ${\bf n}_p(\alpha_{p}, \beta_{p})$ and roll or spin angle $0 \le \gamma_p \le \frac{2 \pi}{s}$ in Figure \ref{fig:spin}. The initial rotation direction is given by the initial relativistic eigenspinors  ${\bf \chi}_{po}^{\uparrow, \downarrow}$. The particles are rotationally symmetric of order $\frac{2 \pi}{s} \ $, with $s \in \mathbb{N}^\star$. Electrons and positrons correspond to $s=1$. 

The $ \ \mathbb{J} =  \{  \gamma_{po} \in [0, \frac{2 \pi}{s}] + \gamma_p  \} \ \times \ \{\uparrow, \downarrow \}$ - valued quaternion least action (\ref{eq:anticommuteRel}) of each particle $p$ in Theorem \ref{th:Hamilton} is
\begin{equation}
\frac{1}{\hbar} {\bf \phi}_{pj}({\bf n}_{p}, \bar{\bf p}_{p}, \gamma_p) =   {\bf \Gamma} \cdot \bar{\bf p}_{p}   \left\{ 
\begin{array}{ll}
\frac{s}{2} (\gamma_p - \gamma_{po})  & \ \text{for \ }  \uparrow \\
\frac{s}{2} (\frac{2 \pi}{s}  - (\gamma_p - \gamma_{po})) & \ \text{for \ }  \downarrow 
\end{array}
\right. \in \mathbb{H} \label{eq:quatactionrel}
\end{equation}
where we exploited the commutation properties of the Dirac matrices (\ref{eq:Gamma}). The actions correspond to the left and right rotations in figure \ref{fig:spin}.

Since $\Delta_M {\bf \phi}_{pj} = {\bf 0}$ in (\ref{eq:density}), an initial classical Dirac spinor distribution remains a Dirac density distribution  for all $t \ge 0$, i.e., we can define the spinor density as $4 i \sqrt{\rho_{p j}}^{\epsilon}  = {\bf \xi}_{p}^{\epsilon} = {\bf \xi}_{p o}^{\epsilon}$ with $ \epsilon \in \mathbb{E}_p = \{ \uparrow, \downarrow \}$. Thus, from  Theorem \ref{th:quantum} and (\ref{eq:sigman}), the relativistic wave spinor (\ref{eq:xi}) of particle $p$,
\begin{equation}
{\bf \psi}_{p}^{\epsilon} ({\bf n}_{p}, \bar{\bf p}_{p}, {\bf \xi}_{p o }^{\epsilon}) =  \sum_{j \in \mathbb{J}}  \sqrt{\rho_{p j}}  e^{\frac{i }{\hbar} {\bf \phi}_{p j}}  = \frac{1}{2} \int_{\gamma_{po} = \gamma_p }^{\gamma_p + \frac{2 \pi}{s}} \sin{\frac{s}{2} (\gamma_p - \gamma_{po})}  d \gamma_{po}  {\bf \Gamma} \cdot \bar{\bf p}_{p} \ {\bf \xi}_{po}^{\epsilon} = {\bf \Gamma} \cdot \bar{\bf p}_{p} {\bf \xi}_{p o }^{\epsilon} \ \ \ \ \ \ \label{eq:psigamma}
\end{equation} 
solves the Dirac equation (\ref{eq:DiracEq}) with every summand $\sqrt{\rho_{p j}}  \ e^{\frac{i }{\hbar} {\bf \phi}_{p j}}$. It computes the wave spinor ${\bf \psi}_{p}^{\epsilon}$  as the geometric projection of an initial relativistic spinor ${\bf \xi}_{po}^{\epsilon}$ on a filter of direction ${\bf n}_p$ or  as a solution of the relativistic eigenspinor equation (\ref{eq:chidownrel}) for ${\bf \psi}_{p}^{\epsilon} = {\bf \xi}_{p}^{\epsilon} = {\bf \xi}_{po}^{\epsilon}$ in free space. (\ref{eq:psigamma}) holds for a electron with positive constant energy $E_{p+}$ and a positron with constant negative energy $E_{p-}$ both with $s=1$. 

Consider now a positron with known relativistic momentum $\bar{\bf p}_{-}$ in (\ref{eq:relativity}) and negative energy $E_- \ $. Its  eigenspinor ${\bf \xi}_{\uparrow}^-$ of (\ref{eq:xi}) solves (\ref{eq:psigamma}), and we consider the positron path traveling against time \cite{feynman_quantum_1998}. A plastic collision of the positron with a photon of momentum $\bar{\bf p}$ from section \ref{Maxwell} leads to a new relativistic momentum $\ \bar{\bf p}_{+} = \bar{\bf p} - \bar{\bf p}_{-} \ $. This in turn leads to an electron with known relativistic momentum $\bar{\bf p}_{+}$ in (\ref{eq:relativity}), whose energy $E_+$ is positive if the energy of the photon was larger than $E_o - E_- \ $, with $E_o =M_o c^2$ the rest energy. The eigenspinor of the electron ${\bf \xi}_{\uparrow}^+$ of (\ref{eq:xi}) now solves (\ref{eq:psigamma}).

Although the above creation of electrons and positrons is a well-established Feynman diagram \cite{feynman_quantum_1998, Weinberg1964} in QED, the novelty is that it is now purely derived from the relativistic action of quaternions (\ref{eq:H1Rel}, \ref{eq:quatactionrel}). Since Theorem \ref{th:quantum} applies to multiple particles, a second quantization is not needed to compute particle creation or annihilation phenomena.
}{Creation}

\section{Concluding Remarks}\label{concluding}

This paper shows that the Schr\"odinger equation and its relativistic counterparts can be solved exactly from a $\mathbb{J}$-valued action of Theorem \ref{th:Hamilton} and the classical densities along the associated multipaths. Identical wave functions and experimentally observed probability distributions can thus be obtained from three different interpretations:
\begin{itemize}
    \item The Schr\"odinger (\ref{eq:Schroed}), Klein-Gordon (\ref{eq:KleinGordon}), Pauli (\ref{eq:Pauli}), Dirac (\ref{eq:DiracEq}), and Maxwell (\ref{eq:Maxwell}) equations, which have no particle path until the wave function collapses at a measurement.
    \item The non-relativistic Feynman path integral (\ref{eq:Feynman}), which has an $\infty^{\infty}$ of time-sliced zig-zag paths with non-classical actions~\cite{Feynman48}. 
    \item Theorem \ref{th:quantum}, which uses the $\mathbb{J}$-valued classical multipaths or actions from Theorem \ref{th:Hamilton} as a subset of all zig-zag paths in Feynman's path integral (\ref{eq:Feynman}). The quantum probability matrix $\varrho$ is generated from the initial classical density distribution $\rho$ propagated along all classical determined paths of each action branch. 
    
    Here the wave collapse at a measurement stems from the collapse of a classical density distribution to a Dirac impulse when it passes through a measurement device in Lemma \ref{lem:measurement}.  
    At measurement time, the initial condition in ${\bf x}_o$ and ${\bf p}_o$ of Theorem \ref{th:Hamilton} is completed by the measurement ${\bf y}_k \ $, which determines in Theorem \ref{th:Hamilton} the $\mathbb{J}$ classical multipaths (\ref{eq:branchpointset})  from $\{ {\bf x}_o \ \veebar \ {\bf p}_o \}$ to ${\bf y}_k \ $.
    
\end{itemize}
The fundamental quantum postulates on the existence of a wave function, its propagation with the Schr\"odinger equation in Theorem \ref{th:quantum} and the wave collapse at a measurement in Lemma \ref{lem:measurement} are derived from the classical Theorem \ref{th:Hamilton}. Furthermore, analytic computations of the classical action are simpler than solving the Feynman path integral and potentially easier than solving the Schr\"odinger equation directly. In addition, Theorem \ref{th:quantum} is a multi-particle result. 

The $\mathbb{J}$ classical multipaths in Theorem \ref{th:quantum} and Lemma \ref{lem:measurement} are strictly determined by the initial and final conditions. 
In the double slit experiment, the probabilistic quantum observation results from the non-Lipschitz constraint force in the slit. For the harmonic oscillator, the Coulomb wave, the particle in the box, or the spinning particle, the initial probabilistic density distribution is classically propagated forward in time. In the EPR experiment~\cite{Aspect,clauser1969}, Theorem \ref{th:Hamilton} determines a constant angular momentum ${\bf \chi}^{\uparrow}_{o}, {\bf \chi}^{\downarrow}_{o} \ $ over time, and Lemma \ref{lem:measurement} in turn allows a classical interpretation that the decision which spin direction is sensed behind the filters is already taken when the particles separate.


Theorems \ref{th:Hamilton} and \ref{th:quantum} yield exact and purely classically-based derivations of well-known quantum wave functions, e.g., for the two-slit experiment (\Ex{Twoslit}), the Aharonov-Bohm effect (\Ex{Aharonov}), a  particle in a box (\Ex{box}), 
quantum tunneling (\Ex{barrier}),  the hydrogen atom (\Ex{Coulomb}), a spinning particle and entanglement (\Ex{spin}), and relativistic electron-positron creation (\Ex{Creation}).  


Motivated by \Ex{barrier}, current research focuses on systematically deriving complex actions for general nonlinear potentials. This may enable exact wave computations where so far perturbation theory \cite{Heisenberg1925} had to be used.
The action-based perspective may also simplify the study of systems modeled as composites of quantum and classical dynamics~\cite{Hooft2007,Dalibard1993MCWF,WangMakri2019}. 
Exact quantum simulations may be obtained from classical physical systems with matching least action paths. 
The differentiability of the classical paths may make machine learning techniques readily applicable, e.g., in computational quantum chemistry.
Finally, the ability to derive quantum quantities just from the $\mathbb{J}$-valued classical action paths may have implications on some of the assumptions in quantum information processing.

\noindent {\bf Acknowledgements} \ \ This paper benefited from discussions with Alain Aspect, Michel Devoret, Steven Girvin, Serge Haroche, Christian Pehle, Carlo Rovelli, Stephan Schiller, and Pierre Rouchon. We also thank the anonymous reviewers for important suggestions.

\normalem
\bibliographystyle{abbrv}
{\fontsize{10}{0} \selectfont \bibliography{References}{}}

@article{Aspect,
author = {Aspect, A.},
journal = {Nature},
title = {{Bell's inequality test more ideal than ever}},
year = {1999}
}

@article{BerryMount1972Semiclassical,
  author       = {Berry, M. V. and Mount, K. E.},
  title        = {Semiclassical Approximations in Wave Mechanics},
  journal      = {Reports on Progress in Physics},
  year         = {1972},
  volume       = {35},
  pages        = {315--397},
  doi          = {10.1088/0034-4885/35/1/306},
}

@article{Aharonov1959,
  author    = {Y. Aharonov and D. Bohm},
  title     = {Significance of Electromagnetic Potentials in the Quantum Theory},
  journal   = {Physical Review},
  year      = {1959},
  volume    = {115},
  number    = {3},
  pages     = {485--491},
  month     = {August},
  issn      = {0031-899X},
  doi       = {10.1103/PhysRev.115.485}
}

@article{Bell1964,
  author = {Bell, J. S.},
  title = {On the {Einstein Podolsky Rosen} paradox},
  journal = {Physics},
  volume = {1},
  number = {3},
  pages = {195--200},
  year = {1964}
}

@article{Bohm1952,
  author    = {David Bohm},
  title     = {A Suggested Interpretation of the Quantum Theory in Terms of "Hidden" Variables},
  journal   = {Physical Review},
  volume    = {85},
  number    = {2},
  year      = {1952}
}

@article{Born1926ZurQD,
  author  = {Max Born},
  title   = {{Zur Quantenmechanik der Sto{\ss}vorg{\"a}nge}},
  journal = {Zeitschrift f{\"u}r Physik},
  year    = {1926},
  volume  = {37},
  pages   = {863--867},
  url     = {https://api.semanticscholar.org/CorpusID:119896026},
}

@article{Bohr1913,
  author    = {Niels Bohr},
  title     = {On the Constitution of Atoms and Molecules},
  journal   = {Philosophical Magazine},
  series    = {Series 6},
  volume    = {26},
  number    = {151},
  pages     = {1--25},
  year      = {1913},
  doi       = {10.1080/14786441308634955}
}

@article{bohr1928quantenpostulat,
  author = {Bohr, Niels},
  title = {{Das Quantenpostulat und die neuere Entwicklung der Atomistik}},
  journal = {Naturwissenschaften},
  volume = {16},
  number = {15},
  pages = {245--257},
  year = {1928},
  doi = {10.1007/BF01504968}
}

@article{Bloch1946NuclearInduction,
  author  = {Bloch, Felix},
  title   = {Nuclear Induction},
  journal = {Physical Review},
  volume  = {70},
  number  = {7--8},
  pages   = {460--474},
  year    = {1946},
  doi     = {10.1103/PhysRev.70.460}
}

@book{zeilinger2000physics,
  title        = {The Physics of Quantum Information: Quantum Cryptography, Quantum Teleportation, Quantum Computation},
  author       = {Bouwmeester, Dirk and Ekert, Artur and Zeilinger, Anton},
  year         = {2000},
  publisher    = {Springer},
  address      = {Berlin, Heidelberg},
  isbn         = {978-3-540-66778-3},
}

@article{clauser1969,
  author  = "John F. Clauser and Michael A. Horne and Abner Shimony and Richard A. Holt",
  title   = "Proposed Experiment to Test Local Hidden-Variable Theories",
  journal = "Physical Review Letters",
  year    = 1969,
  volume  = "23",
  number  = "15",
  pages   = "880--884",
  doi     = "10.1103/PhysRevLett.23.880"
}

@book{Cohen-Tannoudji,
    title = {{Quantum Mechanics}},
    author = {Cohen-Tannoudji, C. and Diu, B. and Laloe, F.},
    year = {2019},
    publisher = {Wiley},
    edition = {Second}
}

@article{Dalibard1993MCWF,
  title = {Monte-Carlo wave function method in quantum optics},
  author = {Dalibard, J. and Castin, Y. and Mølmer K.},
  journal = {Journal of the Optical Society of America B},
  volume = {10},
  number = {3},
  pages = {524--538},
  year = {1993}
}

@article{Devoret1985,
  author = "M. H. Devoret and J. M. Martinis and J. Clarke",
  title = "Measurements of macroscopic quantum tunneling out of the zero-voltage state of a current-biased Josephson junction",
  journal = "Phys. Rev. Lett.",
  volume = "55",
  pages = "1908--1911",
  year = "1985",
  doi = "10.1103/PhysRevLett.55.1908"
}

@article{Dirac28,
  author = {Dirac, P. A. M.},
  title = {The quantum theory of the electron},
  journal = {Proceedings of the Royal Society A},
  volume = {117},
  number = {778},
  pages = {610--624},
  year = {1928}
}

@article{Dirac33,
author = {Dirac, P.},
journal = {Physical J. Soviet Union},
title = {{The Lagrangian in Quantum Mechanics}},
pages = {64--72},
year = {1933}
}

@article{Duru,
author = {Duru, I.H. and Kleinert, H.},
journal = {{Fortschritte Phys.}},
title = {{Quantum Mechanics of H-atoms from path integrals}},
year = {1982}
}

@article{Einstein:1905,
author = {Einstein, Albert},
title = {{Zur Elektrodynamik bewegter Körper}},
journal = {Annalen der Physik},
volume = {322},
number = {10},
pages = {891--921},
year = {1905},
doi = {10.1002/andp.19053221004},
url = {https://de.wikipedia.org/wiki/Spezielle_Relativit%C3%A4tstheorie}
}

@article{Einstein,
author = {Einstein, A.},
journal = {Annalen der Physik},
title = {{Die Grundlage der allgemeinen Relativitätstheorie}},
year = {1916}
}

@article{Einstein-Podolsky,
author = {Einstein, A. and Podolsky, B. and Rosen, N.},
journal = {Physical Review},
title = {Can the quantum-mechanical description of physical reality be considered complete?},
year = {1935}
}

@article{Euler,
    author = {Euler, L.},
    journal = {Académie Royale des Sciences et des Belles Lettres},
    title = {{Principes généraux du mouvement des fluides}},
    year = {1755}
}

@article{Feynman48,
author = {Feynman, R.P. },
journal = {Rev. Modern Physics},
title = {{Space-Time Approach to Non-Relativistic Quantum Mechanics}},
year = {1948}
}

@book{Feynman,
    title = {{Quantum Mechanics and Path Integrals}},
    author = {Feynman, R.P. and Hibbs, A.R.},
    year = {1965},
    publisher = {McGraw-Hill}
}

@book{feynman_quantum_1998,
  address = {Reading, MA},
  author = {Feynman, Richard P.},
  isbn = {9780201360752},
  publisher = {Addison-Wesley},
  title = {Quantum Electrodynamics},
  year = {1998}
}

@article{Fujikawa,
author = {Fujikawa, K.},
journal = {Nuclear Physics},
title = {{Path integral of the hydrogen atom, the Jacobi’s principle of least action and one-dimensional quantum gravity}},
year = {1997}
}

@book{Fulling,
    title = {{Aspects of Quantum Field Theory in Curved Space–Time}},
    author = {Fulling, S.A.},
    year = {1996},
    publisher = {Cambridge Univ. Press}
}

@article{Fraunhofer1823,
  author    = {Joseph von Fraunhofer},
  title     = {{Neue Beobachtungen über die Beugung des Lichts}},
  journal   = {Annal. Physik},
  volume    = {74},
  year      = {1823}
}

@article{Gordon,
    author = {Gordon, W.},
    journal = {{Zeitschrift f\"ur Physik}},
    title = {{Der Compton Effekt nach der Schr\"odingerschen Theorie}},
    year = {1926}
}

@book{Goldstein,
    title = {{Classical Mechanics}},
    author = {Goldstein, H.},
    year = {1980},
    publisher = {Addison-Wesley}
}

@article{Groenewold1946,
  author    = {H. J. Groenewold},
  title     = {On the principles of elementary quantum mechanics},
  journal   = {Physica},
  volume    = {12},
  pages     = {405--460},
  year      = {1946}
}

@article{Heisenberg1925,
  author = {W. Heisenberg},
  title = {{Über quantentheoretische Umdeutung kinematischer und mechanischer Beziehungen}},
  journal = {Zeitschrift für Physik},
  volume = {33},
  pages = {879--893},
  year = {1925},
  doi = {10.1007/BF01328377}
}

@inproceedings{Hamilton,
  title        = {{Second essay on a general method in dynamics}},
  author       = {Hamilton, R.W.},
  year         = 1835,
  booktitle    = {Philosophical Transactions of the Royal Society}
}

@article{hamilton1828rays,
  author    = {William Rowan Hamilton},
  title     = {Theory of Systems of Rays},
  journal   = {Transactions of the Royal Irish Academy},
  volume    = {15},
  pages     = {69--174},
  year      = {1828}
}

@book{heisenberg1958physics,
  author    = {Werner Heisenberg},
  title     = {Physics and Philosophy: The Revolution in Modern Science},
  year      = {1958},
  publisher = {George Allen and Unwin},
  address   = {London},
}

@article{hilbert1904,
  author = {Hilbert, David},
  title = {{Grundzüge einer allgemeinen Theorie der linearen Integralgleichungen}},
  journal = {Nachrichten von der Gesellschaft der Wissenschaften zu Göttingen, Mathematisch-Physikalische Klasse},
  year = {1904},
  url = {http://www.cs.umd.edu/~stewart/FHS.pdf}
}

@article{Hooft2007,
  author    = "{Hooft, Gerard 't}",
  title     = "A mathematical theory for deterministic quantum mechanics",
  journal   = "Journal of Physics: Conference Series",
  volume    = "67",
  number    = "1",
  pages     = "012015",
  year      = "2007"
}

@article{Hund1927tunneling,
  author = {Friedrich Hund},
  title = {{Zur Deutung der Molekülspektren. I. Zeitschr. f. Physik}},
  journal = {Zeitschrift für Physik},
  year = {1927},
  volume = {40},
  url = {https://link.springer.com/article/10.1007/BF01397202}
}

@article{Hawking1975,
  author    = {S. W. Hawking},
  title     = {Particle creation by black holes},
  journal   = {Comm. Mathematical Physics},
  volume    = {43},
  number    = {3},
  year      = {1975}
}

@article{Jacobi,
    author = {Jacobi, C.G.J.},
    journal = {{Journal für die reine und angewandte Mathematik}},
    title = {{\"Uber die Integration der partiellen Differentialgleichungen erster Ordnung}},
    year = {1827}
}

@article{KLein,
    author = {Klein, O.},
    journal = {{Zeitschrift f\"ur Physik}},
    title = {{Quantentheorie und f\"unfdimensionale Relativittstheorie}},
    year = {1926}
}

@book{Kleinert2009,
    title = {{Path Integrals in Quantum Mechanics, Statistics, Polymer Physics and Financial Markets}},
    author = {Kleinert, H.},
    year = {2009},
    publisher = {World Scientific}
}

@book{laplace1799mecanique,
  author    = {Pierre-Simon Laplace},
  title     = {Mécanique Céleste},
  year      = {1799},
  publisher = {J.B.M. Duprat et al.},
  address   = {Paris},
  note      = {Five volumes,1799--1825}
}

@book{beltrami1902ricerche,
  author    = {Eugenio Beltrami},
  title     = {Ricerche di analisi applicata alla geometria},
  year      = {1902},
  publisher = {Vol. 1, Opere Matematiche, Milano},
  pages     = {107--198}
}

@article{lohmiller2005contraction,
  title = {Contraction Analysis of Nonlinear Distributed Systems},
  author = {Lohmiller, Winfried and Slotine, Jean-Jacques},
  journal = {International Journal of Control},
  year = {2005},
  volume = {78},
  number = {9},
  pages = {678--688},
  doi = {10.1080/00207170500130952}
}

@book{Lagrange,
    title = {{Mécanique analytique}},
    author = {Lagrange, J.L.},
    year = {1788},
    publisher = {Chez la veuve Desaint à Paris}
}

@BOOK{landau_quantum,
  AUTHOR = "L. D. Landau and E. M. Lifshitz",
  TITLE = "Quantum Mechanics: Non-Relativistic Theory",
  SERIES = "Course of Theoretical Physics",
  VOLUME = "3",
  PUBLISHER = "Pergamon Press",
  YEAR = "1991",
  ADDRESS = "England"
}

@book{Liboff,
    title = {{Introductory Quantum Mechanics}},
    author = {Liboff, R.L.},
    year = {2002},
    publisher = {Addison Wesley}
}

@article{Lipschitz1869,
  author = {R. Lipschitz},
  title = {{Über die Existenz einer Lösung von Differentialgleichungen}},
  journal = {Mathematische Annalen},
  volume = {1},
  pages = {28--32},
  year = {1869}
}

@article{lohmiller1998contraction,
  title={On contraction analysis for nonlinear systems},
  author={Lohmiller, Winfried and Slotine, Jean-Jacques},
  journal={Automatica},
  volume={34},
  number={6},
  year={1998},
  publisher={Elsevier},
  doi={10.1016/S0005-1098(98)00019-3}
}

@book{Lovelock,
    title = {{Tensors, Differential Forms, and Variational Principles}},
    author = {Lovelock, D. and Rund, H.},
    year = {1989},
    publisher = {Dover}
}

@article{Lorenz1867,
  author    = {L. Lorenz},
  title     = {On the Identity of the Vibrations of Light with Electrical Currents},
  journal   = {Philosophical Magazine},
  series    = {4},
  volume    = {34},
  number    = {230},
  pages     = {287--301},
  year      = {1867},
  url       = {https://en.wikipedia.org/wiki/Philosophical_Magazine}
}

@article{Lorentz1904,
  author = {Hendrik Antoon Lorentz},
  title = {Electromagnetic phenomena in a system moving with any velocity smaller than that of light},
  journal = {Proceedings of the Royal Netherlands Academy of Arts and Sciences},
  volume = {6},
  pages = {809--831},
  year = {1904},
  url = {https://en.wikisource.org/wiki/Electromagnetic_phenomena}
}

@article{Madelung1927,
  author    = {E. Madelung},
  title     = {{Quantentheorie in hydrodynamischer Form}},
  journal   = {Zeitschrift f{\"u}r Physik},
  volume    = {40},
  number    = {3-4},
  year      = {1927},
  doi       = {10.1007/BF01400372}
}

@article{Maxwell1865,
  author = {J. C. Maxwell},
  title = {A Dynamical Theory of the Electromagnetic Field},
  journal = {Philosophical Transactions of the Royal Society of London},
  year = {1865},
  month = {January}
}

@book{Maslov1972,
  author = {Victor P. Maslov},
  title = {Théorie des perturbations et méthodes asymptotiques},
  publisher = {Éditions Mir},
  year = {1972},
}

@article{moyal1949quantum,
  author    = {J. E. Moyal},
  title     = {Quantum mechanics as a statistical theory},
  journal   = {Proceedings of the Cambridge Philosophical Society},
  volume    = {45},
  pages     = {99--124},
  year      = {1949}
}

@book{Newton1687,
  author    = {Isaac Newton},
  title     = {Philosophi{\ae} Naturalis Principia Mathematica},
  publisher = {Jussu Societatis Regiae ac Typis Josephi Streater. Prostat apud plures bibliopolas},
  year      = {1687},
  language  = {lat},
  address   = {London}
}

@article{Pauli,
    author = {Pauli, W.},
    journal = {{Zeitschrift für Physik}},
    title = {{Zur Quantenmechanik des magnetischen Elektron}},
    year = {1927}
}

@article{planck1901theorie,
  title={{Zur Theorie des Gesetzes der Energieverteilung im Normalspektrum}},
  author={Planck, Max},
  journal={Verhandlungen der Deutschen Physikalischen Gesellschaft},
  volume={2},
  pages={237--245},
  year={1901}
}

@article{Pelster,
    author = {Pelster, A. and Wunderlin, A.},
    journal = {{Zeitschrift f\"ur Physik and Condensed Matter}},
    title = {{On the generalization of the Duru-Kleinert-propagator transformations}},
    year = {1992}
}

@article{Picard1890,
  author    = {\'Emile Picard},
  title     = {Sur l’application des méthodes d’approximation successive à l’étude de certaines équations différentielles ordinaires},
  journal   = {Journal de Mathématiques Pures et Appliquées},
  series    = {4},
  volume    = {6},
  year      = {1890},
  pages     = {145--210}
}

@article{Sakoda,
    author = {Sakoda, S.},
    journal = {{Journal of Mathematics and Physics}},
    title = {{On the effective potential of Duru-Kleinert path integrals}},
    year = {2017}
}

@article{Schrodinger1926Atom,
author = {Erwin Schrödinger},
title = {An Undulatory Theory of the Mechanics of Atoms and Molecules},
journal = {Physical Review},
year = {1926},
volume = {28},
doi = {10.1103/PhysRev.28.1049}
}

@article{Schleich2013SchrdingerER,
  title = {Schr{\"o}dinger equation revisited},
  author = {Wolfgang P. Schleich and Daniel M. Greenberger and Donald H. Kobe and Marlan O. Scully},
  journal = {P.N.A.S.},
  year = {2013},
  volume = {110},
  number = {14},
  pages = {5374--5379},
  doi = {10.1073/pnas.1302475110},
  url = {https://www.pnas.org/doi/10.1073/pnas.1302475110}
}

@book{Kepler,
    author = {Kepler, J.},
    title = {{Astronomia Nova}},
    year = {1609},
    publisher = {Astronomia Nova}
}

@article{Schrodinger1926,
  author    = {Erwin Schrödinger},
  title     = {{Quantisierung als Eigenwertproblem}},
  journal   = {Annalen der Physik},
  volume    = {79},
  pages     = {361--376},
  year      = {1926},
  doi       = {10.1002/andp.19263840404}
}

@article{VanVleck1928,
  author = {John H. Van Vleck},
  title = {The Correspondence Principle in the Statistical Interpretation of Quantum Mechanics},
  journal = {P.N.A.S},
  volume = {14},
  number = {2},
  pages = {178--188},
  year = {1928},
  doi = {10.1073/pnas.14.2.178},
  url = {https://www.pnas.org/doi/10.1073/pnas.14.2.178},
}

@article{WangMakri2019,
  author = {F. Wang and N. Makri},
  title = {Quantum-classical path integral},
  journal = {J. Chem. Phys.},
  volume = {150},
  number = {18},
  pages = {184102},
  year = {2019},
  doi = {10.1063/1.5091725}
}

@article{Weinberg1964,
  author = {Steven Weinberg},
  title = {Feynman Rules for Any Spin},
  journal = {Physical Review},
  volume = {133},
  number = {5B},
  pages = {B1318--B1332},
  year = {1964},
  doi = {10.1103/PhysRev.133.B1318}
}

@book{zwiebach2022,
  author    = {Barton Zwiebach},
  title     = {Quantum Mechanics},
  publisher = {The MIT Press},
  year      = {2022},
  address   = {Cambridge}
}

\end{document}